\documentclass{aa}

\usepackage{natbib,twoopt}

\usepackage[varg]{txfonts}
\usepackage{graphicx}
\usepackage{amsmath}
\usepackage{amssymb}
\usepackage{lscape}
\usepackage{color}
\usepackage{xcolor}
\usepackage{longtable}
\usepackage{natbib,twoopt}
\usepackage{booktabs}
\usepackage{footmisc}
\usepackage{txfonts}
\usepackage{scalefnt}
\usepackage[flushleft]{threeparttable}
\usepackage{subcaption}

\newcommand{\Zsun}{\hbox{${Z}_{\odot}$}}
\newcommand{\Msun}{\hbox{M$_{\rm{\odot}}$}}
\newcommand{\Msunyr}{\hbox{M$_{\rm{\odot}}\,\rm{yr^{-1}}$}}

\newcommand{\funit}{erg s$^{-1}$ cm$^{-2}$ }
\newcommand{\ozt}{$1 < z < 2$}
\newcommand{\logM}{$\rm{log(M_*/M_{\odot})}$}
\newcommand{\logOH}{$\rm{12\,+\,log(O/H)}$}

\newcommand{\beagle}{\textsc{beagle}}
\newcommand{\tauSFR}{\hbox{$\tau_\mathrm{SFR}$}}

\newcommand{\OII}{[\mbox{O\,{\sc ii}}]}

\newcommand{\Ha}{H$\alpha$}
\newcommand{\Hb}{H$\beta$}
\newcommand{\Hg}{H$\gamma$}

\begin{document}

\title{Are long gamma-ray bursts biased tracers of star formation? Clues from the host galaxies of the {\it Swift}/BAT6 complete sample of bright LGRBs}
\subtitle{III: Stellar masses, star formation rates and metallicities at $z>1$}
\author{
J.~T.~Palmerio\inst{1,2}\thanks{E-mail: palmerio@iap.fr}
\and S.~D.~Vergani\inst{2,1,3}
\and R.~Salvaterra\inst{4}
\and R.~L.~Sanders\inst{5}
\and J.~Japelj\inst{6}
\and A.~Vidal-Garc\'ia\inst{1,7}
\and P.~D'Avanzo\inst{3}
\and D.~Corre\inst{8}
\and D.~A.~Perley\inst{9}
\and A.~E.~Shapley\inst{5}
\and S.~Boissier\inst{8}
\and J.~Greiner\inst{10}
\and E.~Le Floc'h\inst{11}
\and P.~Wiseman\inst{12,10}
}

\institute{
Sorbonne Universit\'e, CNRS, UMR7095, Institut d'Astrophysique de Paris, F-75014, Paris, France
\and 
GEPI, Observatoire de Paris, PSL University, CNRS, 5 Place Jules
Janssen, F-92190 Meudon, France
\and 
INAF - Osservatorio Astronomico di Brera, via E. Bianchi 46, I-23807 Merate, Italy
\and 
INAF - IASF Milano, via E. Bassini 15, 20133 Milano, Italy
\and
Department of Physics \& Astronomy, University of California, Los Angeles, 430 Portola Plaza, Los Angeles, CA 90095, USA
\and
Anton Pannekoek Institute for Astronomy, University of Amsterdam, Science Park 904, 1098 XH Amsterdam, The Netherlands
\and 
LERMA/LRA, \'Ecole Normale Supérieure, PSL University, Observatoire de Paris, CNRS, Sorbonne Universités, UPMC Univ. Paris 06, F-75005 Paris, France
\and
Aix Marseille Univ, CNRS, CNES, LAM, Marseille, France
\and 
Astrophysics Research Institute, Liverpool John Moores University, 146 Brownlow Hill, Liverpool L3 5RF, UK
\and
Max-Planck-Institute für Extraterrestrische Physik (MPE), Giessenbachstrasse 1, 85748 Garching, Germany
\and 
Laboratoire AIM-Paris-Saclay, CEA/DSM/Irfu - CNRS - Université Paris Diderot, CEA-Saclay, F-91191 Gif-sur-Yvette, France
\and
School of Physics and Astronomy, University of Southampton, Southampton, SO17 1BJ, UK
}

\date{Accepted XXX. Received YYY; in original form ZZZ}

\label{firstpage}

\abstract{}
{
Long gamma-ray bursts (LGRB) have been proposed as promising tracers of star formation owing to their association with the core-collapse of massive stars. 
Nonetheless, previous studies we carried out at $z<1$ support the hypothesis that the conditions necessary for the progenitor star to produce an LGRB (e.g. low metallicity), were challenging the use of LGRBs as star-formation tracers, at least at low redshift.
The goal of this work is to characterise the population of host galaxies of LGRBs at \ozt, investigate the conditions in which LGRBs form at these redshifts and assess their use as tracers of star formation.
}
{
We performed a spectro-photometric analysis to determine the stellar mass, star formation rate, specific star formation rate and metallicity of the complete, unbiased host galaxy sample of the {\it Swift}/BAT6 LGRB sample at \ozt.
We compared the distribution of these properties to the ones of typical star-forming galaxies from the MOSDEF and COSMOS2015 Ultra Deep surveys, within the same redshift range.
}
{
We find that, similarly to $z<1$, LGRBs do not directly trace star formation at \ozt, and they tend to avoid high-mass, high-metallicity host galaxies.
We also find evidence for an enhanced fraction of starbursts among the LGRB host sample with respect to the star-forming population of galaxies. Nonetheless we demonstrate that the driving factor ruling the LGRB efficiency is metallicity. The LGRB host distributions can be reconciled with the ones expected from galaxy surveys by imposing a metallicity upper limit of \logOH$\sim$8.55. We can determine upper limits on the fraction of super-solar metallicity LGRB host galaxies of $\sim20$\%, 10\% at $z<1$, $1<z<2$, respectively.
}
{
Metallicity rules the LGRB production efficiency, which is stifled at $Z\gtrsim0.7$\,\Zsun. Under this hypothesis we can expect LGRBs to trace star formation at $z>3$, once the bulk of the star forming galaxy population are characterised by metallicities below this limit. 
The role played by metallicity can be explained by the conditions necessary for the progenitor star to produce an LGRB. The moderately high metallicity threshold found is in agreement with the conditions necessary to rapidly produce a fast-rotating Wolf-Rayet stars in close binary systems, and could be accommodated by single star models under chemically homogeneous mixing with very rapid rotation and weak magnetic coupling. 
}

\titlerunning{Are LGRBs biased tracers of star formation? Clues from the BAT6 sample}
\authorrunning{J. T. Palmerio et al.}
\maketitle

\section{Introduction}\label{sec:intro}

\begin{table*}
\small
\caption{Stellar mass, star formation rate and metallicity for the hosts of the BAT6 LGRB sample at $1<z<2$. 
References are : 1) \citet{Kruhler2015}; 2) this work; 3) \citet{Perley2016}.
}
\centering

\begin{tabular}{lccccccc}
\toprule
Name  & Redshift & $\mathrm{log(M_*/M_{\odot})}$ & SFR ($M_{\odot}/yr$) & 12 + log(O/H) (M08) & $\mathrm{M_{ref}}$   & $\mathrm{SFR_{ref}}$ & $\mathrm{Z_{ref}}$ \\
\midrule
091208B	& 1.0633 	& $<8.3  ^{*    }_{     }$	& $      ^{      }_{      }$ & $     ^{     }_{     }$	& 3	&   &   \\
080413B	& 1.1012 	& $9.5   ^{+0.2 }_{-0.2 }$	& $2.1   ^{+3.1  }_{-1.2  }$ & $8.35 ^{+0.17}_{-0.29}$	& 2	& 1 & 2 \\
090926B	& 1.2427 	& $9.9   ^{+0.1 }_{-0.1 }$	& $12.1  ^{+23.0 }_{-6.5  }$ & $8.48 ^{+0.09}_{-0.16}$	& 2	& 2 & 2 \\
061007 	& 1.2623 	& $8.9   ^{+0.4 }_{-0.5 }$	& $4.4   ^{+6.2  }_{-2.1  }$ & $8.13 ^{+0.11}_{-0.23}$	& 2	& 2 & 2 \\
061121 	& 1.3160 	& $9.4   ^{+0.1 }_{-0.1 }$	& $58.5  ^{+33.8 }_{-17.6 }$ & $8.51 ^{+0.03}_{-0.04}$	& 2	& 2 & 2 \\
071117 	& 1.3293 	& $<9.8  ^{\dagger}_{   }$	& $>2.8  ^{      }_{      }$ & $8.54 ^{+0.13}_{-0.25}$	& 2	& 2 & 2 \\
100615A	& 1.3979 	& $8.6   ^{+0.2 }_{-0.2 }$	& $8.6   ^{+13.9 }_{-4.4  }$ & $8.16 ^{+0.18}_{-0.36}$	& 2	& 1 & 2 \\
050318 	& 1.4436 	& $<8.6  ^{*    }_{     }$	& $      ^{      }_{      }$ & $     ^{     }_{     }$	& 3	&   &   \\
070306 	& 1.4965 	& $9.7   ^{+0.1 }_{-0.1 }$	& $90.6  ^{+49.0 }_{-31.0 }$ & $8.43 ^{+0.03}_{-0.04}$	& 2	& 2 & 2 \\
060306 	& 1.5597 	& $10.4  ^{+0.1 }_{-0.1 }$	& $12.4  ^{+47.0 }_{-7.8  }$ & $8.91 ^{+0.16}_{-0.41}$	& 2	& 2 & 2 \\
080605 	& 1.6408 	& $9.6   ^{+0.1 }_{-0.1 }$	& $42.5  ^{+30.5 }_{-18.2 }$ & $8.47 ^{+0.04}_{-0.04}$	& 2	& 2 & 2 \\
050802 	& 1.7117 	& $9.0   ^{*    }_{     }$	& $>1.6  ^{      }_{      }$ & $     ^{     }_{     }$	& 3	& 2 &   \\
080602 	& 1.8204 	& $9.4   ^{+0.1 }_{-0.1 }$	& $>48   ^{      }_{      }$ & $8.69 ^{+0.12}_{-0.21}$	& 2	& 2 & 2 \\
060908 	& 1.8836 	& $9.2   ^{*    }_{     }$	& $      ^{      }_{      }$ & $     ^{     }_{     }$	& 3	&   &   \\
060814 	& 1.9223 	& $10.0  ^{+0.1 }_{-0.1 }$	& $47.5  ^{+72.5 }_{-15.6 }$ & $8.46 ^{+0.10}_{-0.16}$	& 2	& 2 & 2 \\
\bottomrule

\end{tabular}
\tablefoot{
* These galaxies' stellar mass is computed only from the NIR \textit{Spitzer}/IRAC1 magnitudes or limits (\citealt{Perley2016}; see Sect.~\ref{subsec:prop_M*}).\\
$\dagger$  This galaxy is blended with another  source in the IRAC1 observations and partially in the $K_s$ band, therefore we conservatively report is stellar mass as an upper limit.
}
\label{tab:prop_sample}
\end{table*}

Long duration gamma-ray bursts (LGRBs, prompt emission duration longer than 2s) have been shown to be connected to the end of life of massive stars \citep{Woosley1993,Woosley2006a} from their association with core-collapse supernovae (CCSNe; \citealt{Hjorth2003}).
Due to the short-lived nature of massive stars, LGRBs are thus linked to recent ($\sim$ 10\,Myr) star formation (SF) and it has been suggested that their rate is linked to the global star formation rate (SFR) \citep{Porciani2001}.
Complementary to existing methods such as rest-frame UV measurements, LGRBs offer therefore a promising method of tracing SF up to high redshifts ($z \sim 9$ and beyond, \citealt{Salvaterra2009,Salvaterra2013,Tanvir2009}).
Indeed, in addition to their bright afterglows, even at high redshift \citep{Lamb2000}, the detection of LGRBs in the soft $\gamma$-ray domain of the electromagnetic spectrum is largely unaffected by dust.
Various authors have tried to use LGRBs to estimate the SFR density at high redshift (e.g. \citealt{Kistler2008,Robertson2012}), however these studies used intrinsically biased and incomplete samples.
The importance of using a carefully selected, unbiased and complete sample of LGRBs and their host galaxies has since been recognised and various samples have been designed to address this issue, such as TOUGH \citep{Hjorth2012}, {\it Swift}/BAT6 \citep{Salvaterra2012} and SHOALS \citep{Perley2016b}.

Different studies using the host galaxies of these samples have tried to obtain information on the LGRB efficiency, that is the relation between the LGRB rate and the SFR, fundamental for using LGRBs as tracers of the SFR density. 
Factors that can impact this relation can be related to the conditions needed for the progenitor star to produce an LGRB. 
Metallicity is the most commonly invoked factor, as most single-star progenitor models of LGRBs require low metallicity to expel the hydrogen envelope while keeping enough angular momentum, necessary for the production of the GRB jet (e.g. \citealt{Woosley2006,Yoon2006}). 
Due to the cosmological origin of the majority of LGRBs it is not possible to study directly the progenitor stars, their environment and their remnants. 
Therefore current studies focus on the properties of the LGRB host galaxies to gather information on the LGRB efficiency.
The results obtained to date using complete unbiased samples of LGRB host galaxies \citep{Vergani2015,Perley2016,Japelj2016}, agree on the fact that there is a preference for LGRBs to explode in sub-solar metallicity host galaxies (see also \citealt{Bignone2017} on results using the {\it Illustris} simulation). 
Nonetheless extremely low metallicities are not required, and host galaxies having super-solar metallicities are not excluded (see e.g. \citealt{Savaglio2012a}), even if much rarer than expected from a direct relation between LGRB rate and SFR. 

The results obtained from the studies above are based on the comparison of the properties of LGRB host galaxies with those of representative star-forming galaxies sampled through galaxy surveys. 
Due to the faintness of a considerable fraction of the LGRB host galaxies, to date such a comparison, especially when involving spectroscopically-derived properties (SFR, metallicity), has been performed in detail only at $z<1$ \citep{Kruhler2015,Japelj2016}. 
Improvements of existing photometric surveys (e.g. COSMOS2015, \citealt{Laigle2016}), and the emergence of deep spectroscopic surveys (e.g. VUDS, \citealt{Lefevre2015}) with access to the near-infrared (e.g. MOSFIRE Deep Evolution Field, i.e., MOSDEF survey, \citealt{Kriek2015}) allow us now to investigate the LGRB efficiency 
by comparing the properties of complete samples of LGRB hosts to samples of typical star-forming field galaxies in detail also at $z>1$.

This paper is organised as follows.
In Section~\ref{sec:sample} we present our sample selection, the observations and analysis of our LGRB hosts, and characterise their properties and the evolution of these properties with redshift.
In Section~\ref{sec:compar_samples} we compare our sample with surveys of field galaxies. 
We discuss our results in more detail in Section~\ref{sec:discussion} and our conclusions are summarised in Section~\ref{sec:conclu}.

All errors are reported at 1$\sigma$ confidence unless stated otherwise. We use a standard cosmology \citep{Planck2014}: $\Omega_{\rm m} = 0.315$, $\Omega_{\Lambda} = 0.685$, $H_{0} = 67.3$ km s$^{-1}$ Mpc$^{-1}$. 
The stellar masses (M$_*$) and SFRs are determined using the Chabrier initial mass function (IMF, \citealt{Chabrier2003}).

\section{The {\it Swift}/BAT6 sample of LGRB host galaxies at $z>1$}\label{sec:sample}

\subsection{Selection}\label{subsec:sample_select}

Our sample is composed of the hosts of the \textit{Swift}/BAT6 sample \citep{Salvaterra2012} of bright (peak flux $P_{15-150\,\rm{keV}}\, \geq \,2.6~ \rm{ph\,cm^{-2}\,s^{-1}}$) LGRBs with favourable observing conditions for optical follow-up \citep{Jakobsson2006}. This selection results in 58 LGRBs with a 97\% redshift completeness, extending up to $z \sim 6$.
No correlations have been found between the prompt emission properties (peak energy, luminosity) of LGRBs and their host galaxies' properties (see \citealt{Levesque2010a,Japelj2016}, and Fig~\ref{fig:prompt_correl} of the Appendix for our sample up to $z = 2$).
Therefore, by construction, our sample is statistically representative of the whole LGRB host galaxy population (including dark LGRBs, \citealt{Greiner2011, Melandri2012}).
For the purpose of this paper, we restrict ourselves to the redshift range $1<z<2$ (see Table~\ref{tab:prop_sample}), building on the previous papers of \citet{Vergani2015} and \citet{Japelj2016} that considered the $z<1$ range.

\subsection{Stellar mass}\label{subsec:prop_M*}

To determine the host galaxy stellar masses we used photometric measurements (typically covering the visible to near-infrared wavelength range) from the literature, complemented with new values that we measured from archival data for GRB\,061007, GRB\,100615A and GRB\,090201.
All of the values and references are reported in the appendix (Tables~\ref{tab:photometry_opt} and \ref{tab:photometry_nir}).

We fit the available observational constraints (excluding non-detections\footnote{However we verified that the results from the SED fitting do not violate the limits imposed by the photometric non-detectrions.}) on the emission-line fluxes and broad-band photometry of the galaxies in our sample using the Bayesian spectral interpretation tool \beagle\ (\citealt{Chevallard2016}; version 11.3).
The version of \beagle\ we use relies on the models of \citet{Gutkin2016}, who follow the prescription of \citet{Charlot2001} to describe the emission from stars and the interstellar gas. 
In particular, the models are computed combining the latest version of the \citet{Bruzual2003} stellar population synthesis model with the standard photoionisation code \textsc{cloudy} \citep{Ferland2013}. 
We use three parametrisations for the star formation histories of model galaxies in \beagle\: constant star-formation, an exponentially declining function $\psi(t)\,\propto\,\exp(-t/\tauSFR)$ and an exponentially delayed function $\psi(t)\,\propto\,t\,\exp(-t/\tauSFR)$. 
For the exponentially declining and exponentially delayed functions, we let the star formation timescale and the star-formation freely vary in the ranges $7\leq\log(\tauSFR/\mathrm{yr})\leq11.5$ and $-4\leq\log(SFR/\Msun\mathrm{yr}^{-1})\leq4$. 
Finally, we superpose on the exponentially delayed function a current burst with a variable duration of $6\leq\log(t_{\mathrm{current}}/\mathrm{yr})\leq9$. 
For the three star formation histories, we let the age of the galaxy vary in the range $6.0\leq\log(\mathrm{age}/\mathrm{yr})\leq10.15$ and we adopt a standard \citet{Chabrier2003} initial mass function.
We further adopt the same metallicity for stars and star-forming gas ($Z=Z_{ISM}$) and assume that all stars in a galaxy have the same metallicity, in the range $-2.2\leq\log(Z/Z_\odot)\leq0.25$.
Finally, we let the stellar mass vary in the range $4\leq \log(\rm M_*/M_\odot) \leq 12$.

The stellar mass values reported in Table~\ref{tab:prop_sample} are the median of the probability distribution functions from the best-fitting SFH/attenuation prescription 
which were chosen as having the lowest $\chi^2$ while also having predicted SFRs and metallicities consistent with the ones measured from spectroscopy. The best fit SED for each fitted host are shown in Appendix~\ref{app:sampleBAT6}.

In general, the stellar masses found are consistent within errors independently of the SFH or dust attenuation chosen (the only debated case is GRB\,061121 for which the stellar mass spans values from $7\times10^{8}$ to $2\times10^{10}$\,\Msun, we chose the stellar mass corresponding to the SFH prescription that yields SFR and metallicity values consistent with the ones derived by spectroscopy; we note that using the stellar mass value of \logM$\sim10$ would not change the results of our study). The largest dispersion between the stellar mass values obtained from the different SFH and dust attenuation prescriptions is $\sim 0.5$\,dex.
We cross-checked the stellar mass values with the \textsc{CIGALE} SED code (\citealt{Noll2009}), and values from \cite{Kruehler2017b}, derived using the \textsc{LePhare} SED code \citep{Arnouts1999,Ilbert2006}. 
Even if a detailed analysis on the different SED codes to determine stellar masses is far beyond the scope of this paper, we stress that the stellar mass values found are consistent within the errors, and that the overall results of this study would remain unchanged independently of the choice of the aforementioned codes.

\begin{figure}[!ht]
 \centering
   \includegraphics[width=\hsize]{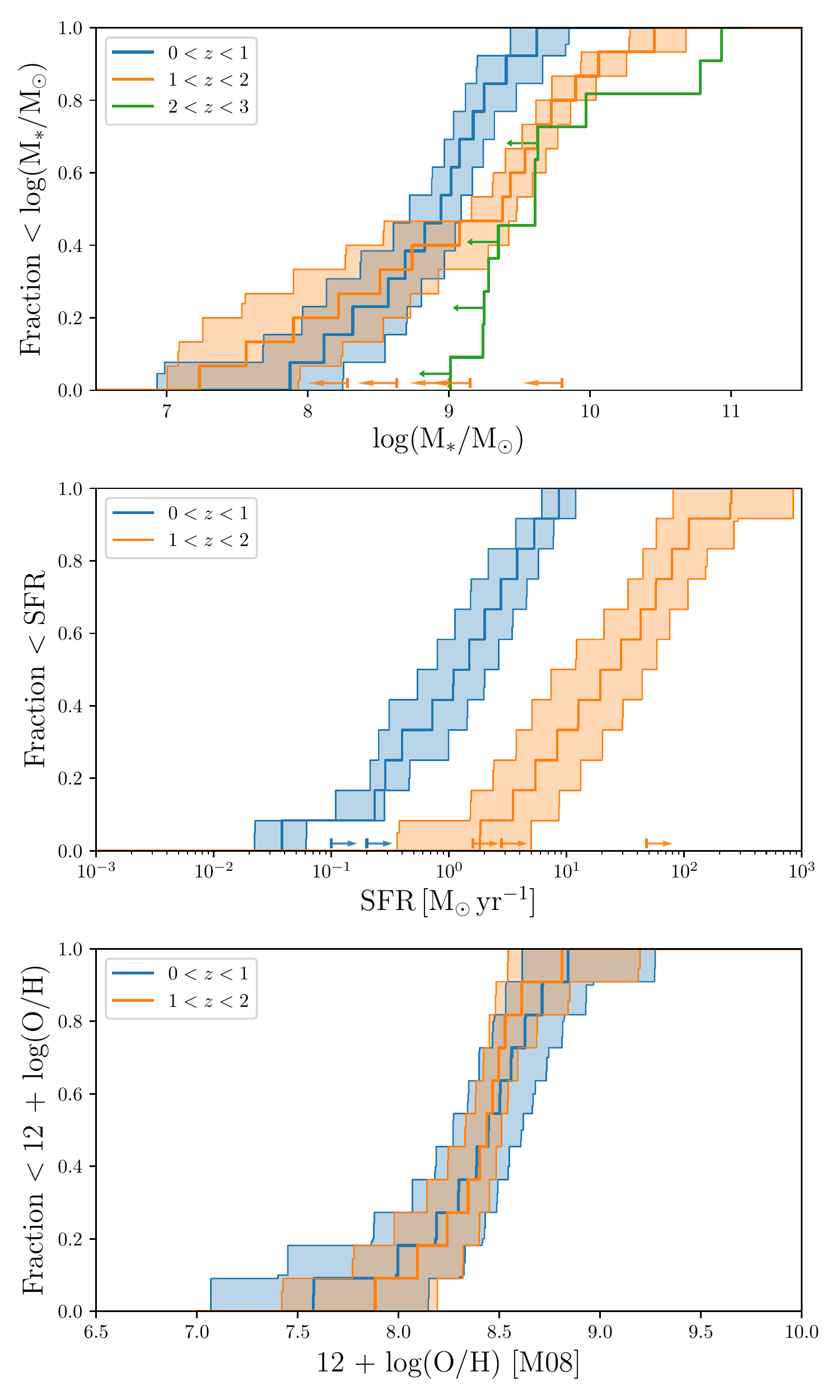}
\caption{
\small
Cumulative distributions of stellar mass (upper panel), SFR (middle panel) and metallicity (bottom panel) for the hosts of the BAT6 sample at different redshift ranges.
Upper and lower limits are represented as arrows at the bottom of the plots.
The shaded area represents the 95\% confidence interval around the CDFs.
The methodology to create these CDFs is presented in Sect.~\ref{subsec:Bayes}}
\label{fig:ECDF}
\end{figure}

We noticed a discrepancy (see also \citealt{Corre2018,Arabsalmani2017,Heintz2017}) when computing stellar masses from SED fitting compared to values based on the rest-frame near-infrared (NIR) magnitude only (e.g. from \citealt{Perley2016}, used also in \citealt{Vergani2017}).
These stellar mass values are mostly overestimated compared to the values derived by SED fitting.
This effect is known, especially at lower stellar masses due to the variations in the mass-to-light ratio as a function of stellar mass \citep{Ilbert2010}.
It should be noted however that \citet{Perley2016} tried to correct for this effect by using a mass-to-light conversion factor that is not simply a linear factor but is a function of $z$ and galaxy luminosity, fitted based on a template model of galaxy evolution.
Due to the lack of wide photometric coverage, 
for 4 of our 15 hosts at \ozt\ (GRB\,091208B, GRB\,050318, GRB\,050802, GRB\,060908) it was not possible to perform a SED fitting, therefore the stellar masses are computed with the method described in \citet{Perley2016}, with the aforementioned caveats. These values are considered as upper limits in the analysis.
However, as explained later (see Sect.~\ref{sec:compar_samples}), they are discarded when performing the statistical test of Sect.~\ref{sec:compar_samples} as they do not comply with the limits of the surveys.   

The resulting stellar mass cumulative distribution for the hosts of the BAT6 sample is shown in Fig~\ref{fig:ECDF}, in the top panel.
There is an evolution towards higher median mass between $z < 1$ and \ozt. As LGRB host galaxies are selected only by the fact that they host an LGRB explosions, and as we are considering an unbiased and complete sample of LGRB host galaxies, the stellar mass evolution we find is not a selection effect and is intrinsic to the properties of LGRB host galaxies. Nevertheless, we anticipate that higher stellar mass values would be expected considering the SFR determined in Sect.~2.3 and the relation found between stellar mass and SFR in SF galaxies (e.g. \citealt{Shivaei2015}).  

We also plot the distribution of the stellar masses or limits for the BAT6 LGRB host galaxies at $2<z<3$ (see Tab.~\ref{tab:prop_sample_2z3}). Those were determined from rest-frame NIR observations only, and (with the exception of GRB\,090201) published by \cite{Perley2016}. 
The distribution at $2 < z < 3$ is riddled with upper limits, and given the different methodology used for the stellar mass determination (and its caveats), we can only tentatively conclude that the median stellar mass does not seem to increase significantly with respect to the one at \ozt.

\subsection{Star Formation Rate and Metallicity}\label{subsec:prop_SFR_logOH}

SFRs and metallicities were determined using the host galaxy spectra.
The data at $z>1$ come from the VLT/X-Shooter spectrograph \citep{Vernet2011}, and the spectra have already been presented in \citet{Kruhler2015} and \citet{Vergani2017}.
The large wavelength coverage (3000 to 25000 \AA) and sensitivity of X-Shooter allow us to detect the strongest rest-frame optical emission lines up to $z = 2$, ensuring a homogeneous methodology for the determination of star formation rates and metallicities.

We performed a new data reduction and analysis of the data, with the standard Esoreflex pipeline (version 2.7.3, \citealt{Modigliani2010}) using the nodding recipe. 
The spatial width of the 2D to 1D spectrum extraction was scaled according to the spatial width of the detected emission lines to maximise the signal to noise ratio.
The flux calibration was cross-checked with the host photometry when available, or otherwise with a telluric standard star taken at similar airmass and seeing, to account for any slit loss or absolute calibration inconsistencies (see \citealt{Japelj2016}).
Emission lines were measured using IRAF\footnote{IRAF is distributed by the National Optical Astronomy Observatories, which are operated by the Association of Universities for Research in Astronomy, Inc., under cooperative agreement with the National Science Foundation.} by fitting a one (or more when relevant) component Gaussian function and cross-checked by comparing to the flux resulting from direct integration under the line profile.
The resulting fluxes are compiled in Table~\ref{tab:line_flux}.
In case of a non-detection, a 3$\sigma$ upper limit is quoted.
The measurements are consistent within the errors with the values reported by \citet{Kruhler2015} and \citet{Vergani2017}.

The measured emission line fluxes were corrected for Galactic extinction using the extinction curve of \citet{Pei1992} and the extinction map of \citet{Schlafly2011}.
The Balmer line fluxes were not corrected for Balmer absorption due to the absence of a detectable continuum in most hosts and its weakness in LGRB hosts as expected from their low stellar masses \citep{Zahid2011}.
The fluxes were also corrected for the host intrinsic extinction, with the A$_V$ measured using the Balmer decrement (assuming case B recombination, \citealt{Osterbrock1989}) and an SMC extinction curve following the findings of for example \citealt{Japelj2015}.

SFRs were determined using the dust-corrected \Ha\, luminosities, following \citet{Kennicutt1998} scaled to the IMF of \cite{Chabrier2003}. 
In the few cases where it was not possible to correct for dust extinction, the SFRs are reported as lower limits.
As shown in Figure~\ref{fig:ECDF}, panel (b), the median SFR increases from $\sim$ 1.3$^{+0.9}_{-0.7}$ \Msunyr\, at $z < 1$, to $\sim$ 24$^{+24}_{-14}$ \Msunyr at \ozt, in agreement with \cite{Kruhler2015}.

\begin{figure*}[!ht]
\begin{subfigure}[b]{0.48\textwidth}
\centering
\includegraphics[width=0.9\textwidth]{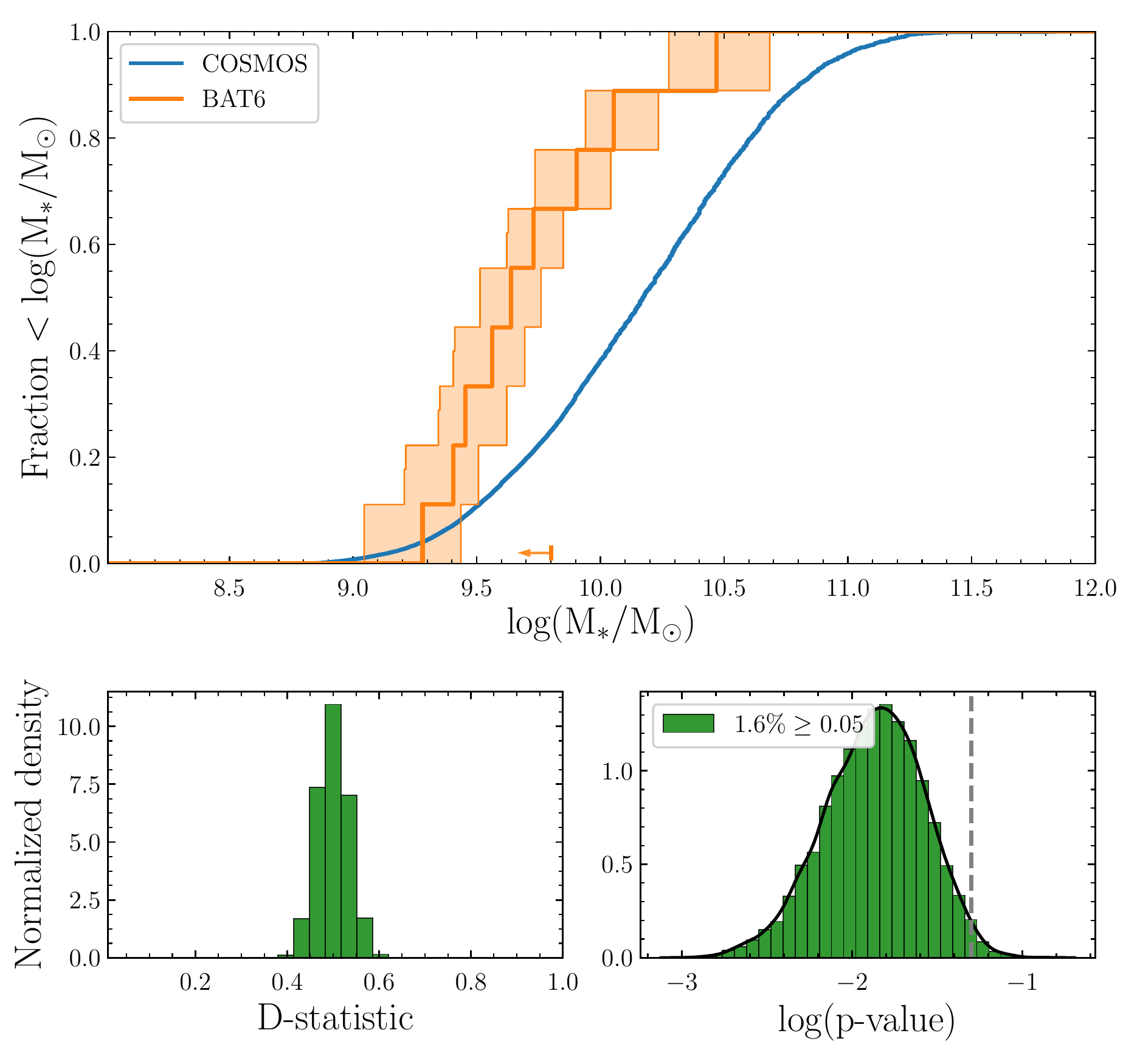}
\end{subfigure}
\hfill
\begin{subfigure}[b]{0.48\textwidth}
\centering
\includegraphics[width=0.9\textwidth]{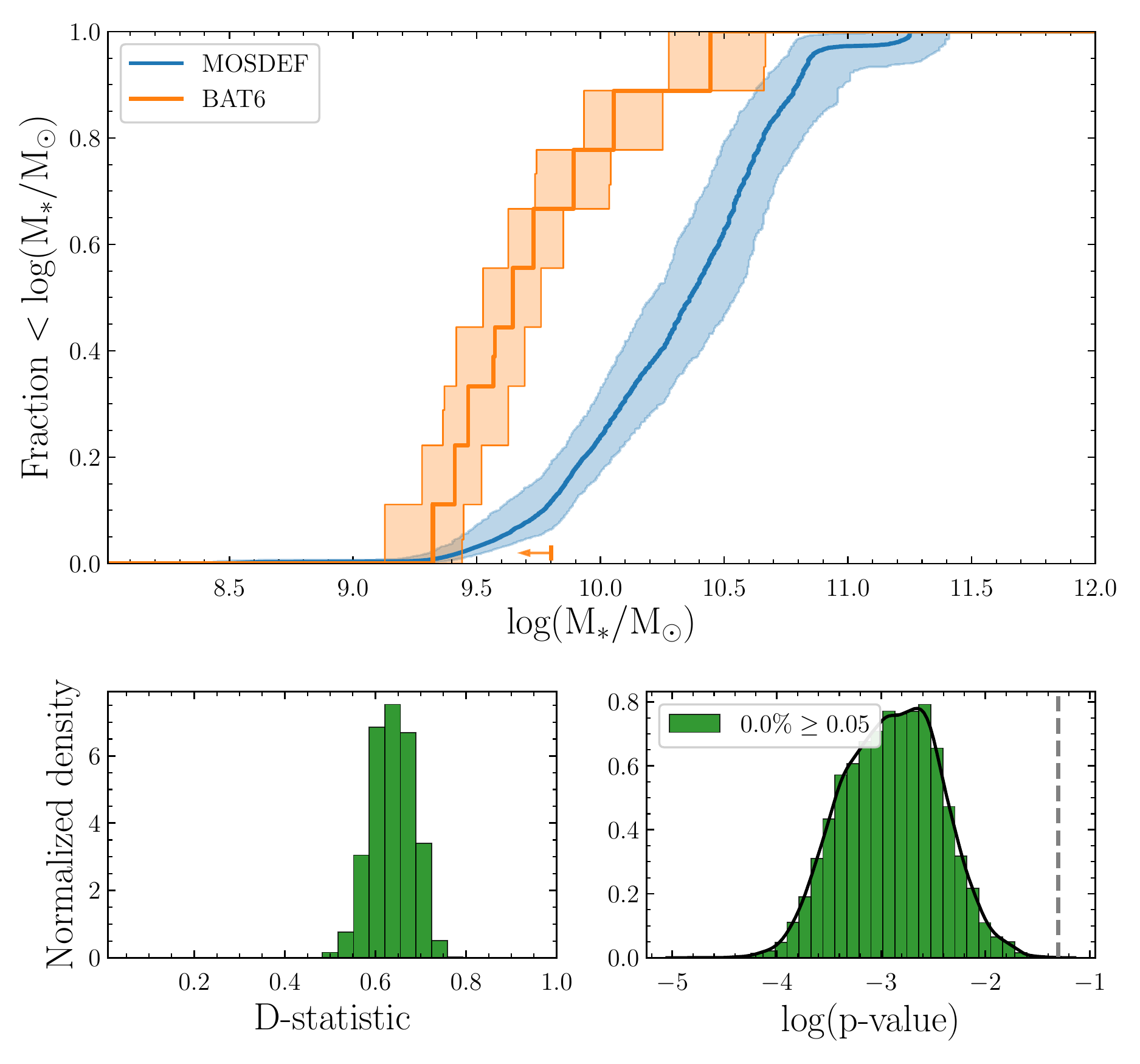}
\end{subfigure}
\caption{
\small
Top panels: Cumulative stellar mass distribution for the hosts of the BAT6 sample (orange) and the star-forming galaxies from the COSMOS2015 Ultra Deep catalogue (blue, left panel) and MOSDEF (blue, right panel) at \ozt.
The COSMOS2015 Ultra Deep and MOSDEF CDFs are weighted by SFR.
Limits are indicated by arrows at the bottom of the plot.
Bottom panels: Normalised histogram of the maximum distance between the BAT6 and the survey CDFs for each Monte Carlo realisation and of the p-value from the two-sample K-S test computed for each Monte Carlo realisation.
The black curve represents the Gaussian kernel density estimation.
The vertical dashed line indicates a p-value of 0.05, above which it is no longer possible to reject the null hypothesis that the two samples are drawn from the same distribution at a 95\% confidence level.
}
\label{fig:mass_CDF}
\end{figure*}

Gas phase metallicities are notoriously hard to determine at high redshift by direct electron temperature methods due to the weakness of the [OIII]$\lambda$4363 line. 
Instead, alternative methods based on the calibration of strong line ratios are commonly used.
Each calibrator has its own relative scale (see \citealt{Kewley2008} for more details).
It is therefore important to use the same method to determine metallicity for all the host galaxy in our sample. 
Here we infer the metallicity from the method developed by \citet{Maiolino2008} (referred to as M08) which relies on multiple calibrators simultaneously, taking advantage of all the emission lines detected. 
The bottom panel of Figure~\ref{fig:ECDF} indicates that, contrary to stellar mass and star formation, the metallicity distribution of LGRB hosts does not seem to evolve (see also \citealt{Kruhler2015}). 
This provides a first clue suggesting that metallicity is a regulatory factor in the production of LGRBs, which is in line with previous studies \citep{Vergani2015,Perley2016}.

\section{Comparison with the star-forming galaxy population}\label{sec:compar_samples}

If we assume that LGRBs are direct tracers of SF, then more SF equates to a higher chance of producing an LGRB (for a fixed stellar IMF).
Hence from a statistical point of view we expect the various distributions of the properties of LGRB hosts to follow the ones of the general population of star-forming galaxies weighted by their SFR. 
The lack of agreement between these distributions can be an indication of a factor regulating the production of LGRBs.
\citet{Vergani2015} and \citet{Japelj2016} have already shown discrepancies between the distributions of the {\it Swift}/BAT6 LGRB hosts properties and the SFR-weighted ones of star-forming galaxies at $z < 1$, (see also \citealt{Kruhler2015}, \citealt{Perley2016} and \citealt{Schulze2015} tackling the same issue using other samples).
Here we aim to extend this analysis to higher redshift.
Owing to a low number of objects and, in some cases, limits or large errors, we employ a Bayesian approach to provide robust statistical estimates, which we describe in Sect~\ref{subsec:Bayes}.

\subsection{Comparison samples}\label{subsec:sample_compare}
\subsubsection{COSMOS 2015 Ultra Deep}

The COSMOS2015 \citep{Laigle2016} is a deep ($\rm{K_s}~\leq~24.7$) photometric survey of half a million galaxies at $z<6$, with wavelength coverage from the near-UV to the infrared. 
Within this catalogue we selected the star-forming galaxies of the COSMOS2015 Ultra Deep stripes (COSMOS2015UD) from the ESO phase 3 archive system\footnote{\label{footnote:ESO3}http://www.eso.org/qi/}.
The advantage of COSMOS2015UD relies in the large number of objects ($\sim10^{4-5}$) with available stellar masses and accurate photometric redshifts.
These stellar masses were determined by SED fitting with the \textsc{LePhare} code using a \citet{Chabrier2003} IMF (see \citealt{Ilbert2015} for more details).
While comparing the properties of the BAT6 LGRB host galaxies with COSMOS2015UD, we take into account its redshift-dependent mass completeness and remove the LGRB hosts with stellar masses below this limit, resulting in a comparison sub-sample of ten hosts at \ozt.
The COSMOS2015UD SFR are dust-corrected and obtained from SED fitting without the Infrared photometry (\url{http://www.eso.org/rm/api/v1/public/releaseDescriptions/100}).

\subsubsection{The MOSDEF survey}
The MOSDEF survey \citep{Kriek2015} is a deep near-infrared spectroscopic survey of galaxies at $1.37\leq z\leq3.80$ that was carried out using the Multi-Object Spectrometer for Infra-Red Exploration (MOSFIRE, \citealt{McLean2012}) on the 10~m Keck~I telescope.
Targets were selected in three redshift ranges ($1.37\leq z \leq1.70$, $2.09\leq z \leq2.61$, and $2.95\leq z \leq3.80$) in which strong rest-frame optical emission lines fall in bands of atmospheric transmission in the near-infrared.
For comparison to the \textit{Swift}/BAT6 LGRB hosts at $1<z<2$, we make use of MOSDEF galaxies in the lowest of these three redshift ranges, at $z\sim1.5$. 
Galaxies were targeted down to fixed rest-optical (observed $H$-band) magnitudes ($H_{\text{AB}}\leq24.0$ at $z\sim1.5$).
We select galaxies with detections of both H$\alpha$ and H$\beta$ at S/N$\geq$3 such that reddening-corrected SFR can be determined.
Requiring detections of both H$\alpha$ and H$\beta$ does not significantly bias the MOSDEF sample above log(M$_*$/M$_{\odot})\sim9.5$ \citep{Shivaei2015,Sanders2018}. AGN were excluded following the prescriptions described in \cite{Shivaei2015} and references therein.
This selection results in a MOSDEF comparison sample of 133 galaxies ranging in redshift from 1.37 to 1.73 with $z_{\text{med}}=1.53$.
 
SFRs were calculated based on reddening-corrected H$\alpha$ luminosity using the \cite{Kennicutt1998} calibration with the \cite{Chabrier2003} IMF, the measured Balmer decrement (H$\alpha$/H$\beta$), and the \cite{Cardelli1989} Milky Way extinction curve.
The MOSDEF stellar masses (see \citealt{Sanders2018}) were estimated by fitting flexible stellar population synthesis models \citep{Conroy2009} to photometry spanning the observed optical to mid-infrared using the SED fitting code FAST \citep{Kriek2009}. 
Solar metallicity, delayed star formation histories, the \cite{Calzetti2000} dust reddening curve, and the \cite{Chabrier2003} IMF were assumed for the SED fitting. 
For comparison with the LGRB host galaxies, SFR(H$\alpha$) and stellar mass values were calculated assuming the \cite{Planck2014} cosmology\footnote{ The papers published previously in the MOSDEF collaboration used a different cosmology.}, the same as for the BAT6 host sample.

The MOSDEF metallicities used in this paper were determined using the M08 method, in the same way as for the hosts of the BAT6 sample (see Sect 2.3). 
Of the 133 galaxies in the MOSDEF comparison sample, 127 have sufficient emission line information to calculate metallicities using the M08 method. 
When comparing this MOSDEF sample with the BAT6 LGRB host galaxies, we excluded from the comparison 6 LGRB hosts with log(M$_*$/M$_{\odot})<~9.3$ because they fall in a stellar mass range in which the MOSDEF sample is significantly incomplete. 
This results in a BAT6 comparison sub-sample of 9 LGRB hosts. 
We note that the SFRs of the LGRB host galaxies in the BAT6 comparison sub-sample fall within the SFR range of the MOSDEF comparison sample.

\subsection{Bayesian framework}\label{subsec:Bayes}
Our calculations rely on the assumption that the probability distribution function (PDF) for our data can be reasonably well described by an asymmetric Gaussian distribution for which the scale parameter is given by the asymmetric errors and the location parameter is given by the value quoted in our table.
For example, the PDF of a quantity $\mu^{+\sigma_p}_{-\sigma_m}$ is given by:
\begin{equation}
\mathrm{PDF}(x)  = A\,
\begin{cases}
\exp(-\frac{(x-\mu)^2}{2\,\sigma_p^2}) & \text{if}~x \geq \mu, \\
\exp(-\frac{(x-\mu)^2}{2\,\sigma_m^2}) & \text{if}~x < \mu,
\end{cases} 
\end{equation}
where $A$ is the normalisation given by: $$ 1 = \int_{-\infty}^{+\infty} \mathrm{PDF}(x)\,dx $$
In the event of upper limits on the stellar mass of our galaxies, we use a uniform distribution (uninformative prior) between \logM\ = 7 and the upper limit for the comparison of BAT6 sample at different redshifts; when comparing with the COSMOS2015UD and MOSDEF surveys, the lower stellar mass limit is set to the mass completeness of the survey\footnote{The mass completeness of the COSMOS2015UD survey varies with redshift; the value used as lower limit is the mass completeness at the redshift of the host.}.
For lower limits on the SFR or the specific SFR (sSFR) (objects for which no extinction could be derived), we use a uniform distribution between the limit and a maximum SFR calculated by assuming an A$_V$ of 4.
We then estimate the median and 95\% confidence bounds on our cumulative distribution functions (CDFs), by computing 10000 Monte Carlo realisations of our data sampling from the aforementioned PDFs, this confidence interval is represented as a shaded area in the figures showing CDFs.
In a similar fashion, we computed 10000 realisations of the K-S test for each individual CDF when comparing with the MOSDEF and COSMOS2015UD samples\footnote{For the COSMOS2015UD sample, the CDF built from the median values reported in the catalogue was used due to the size of the sample which makes this method statistically less relevant and computationally expensive.}.

\subsection{Stellar mass}\label{subsec:survey_compar_mass}

The top left panel of Figure~\ref{fig:mass_CDF} shows the stellar mass cumulative distribution of the hosts of the BAT6 sample compared to the SFR-weighted distribution of the star-forming field galaxies of the COSMOS2015UD at \ozt.
The distribution of D-statistic and p-values from 10000 Monte Carlo realisations of the 2 sample K-S test are shown in the bottom panels, indicating that the vast majority of realisations exclude the null hypothesis that the two samples are drawn from the same distribution at the 95\% confidence level.
It should be noted (see Sect. 3.1.1) that the SFR used to weight the COSMOS2015UD distribution are obtained from SED fitting. 
It has been shown that SFRs determined in such way can be underestimated at SFRs higher than $\sim 50$ \Msunyr (e.g. \citealt{Reddy2015,Lee2015}). This corresponds to $\sim12\%$ of the COSMOS2015UD SF galaxies.
Considering that high SFR values are normally associated with high stellar mass galaxies, this underestimation would have the effect of increasing the discrepancy between the two distributions.
However, we note also that there is a good consistency between the COSMOS2015UD and MOSDEF (see below) SFR-weighted distributions.
These considerations are also valid for the SFR-weighted SFR and sSFR distributions presented in the following sections.

\begin{figure}[!ht]
\begin{center}
\includegraphics[width=\linewidth]{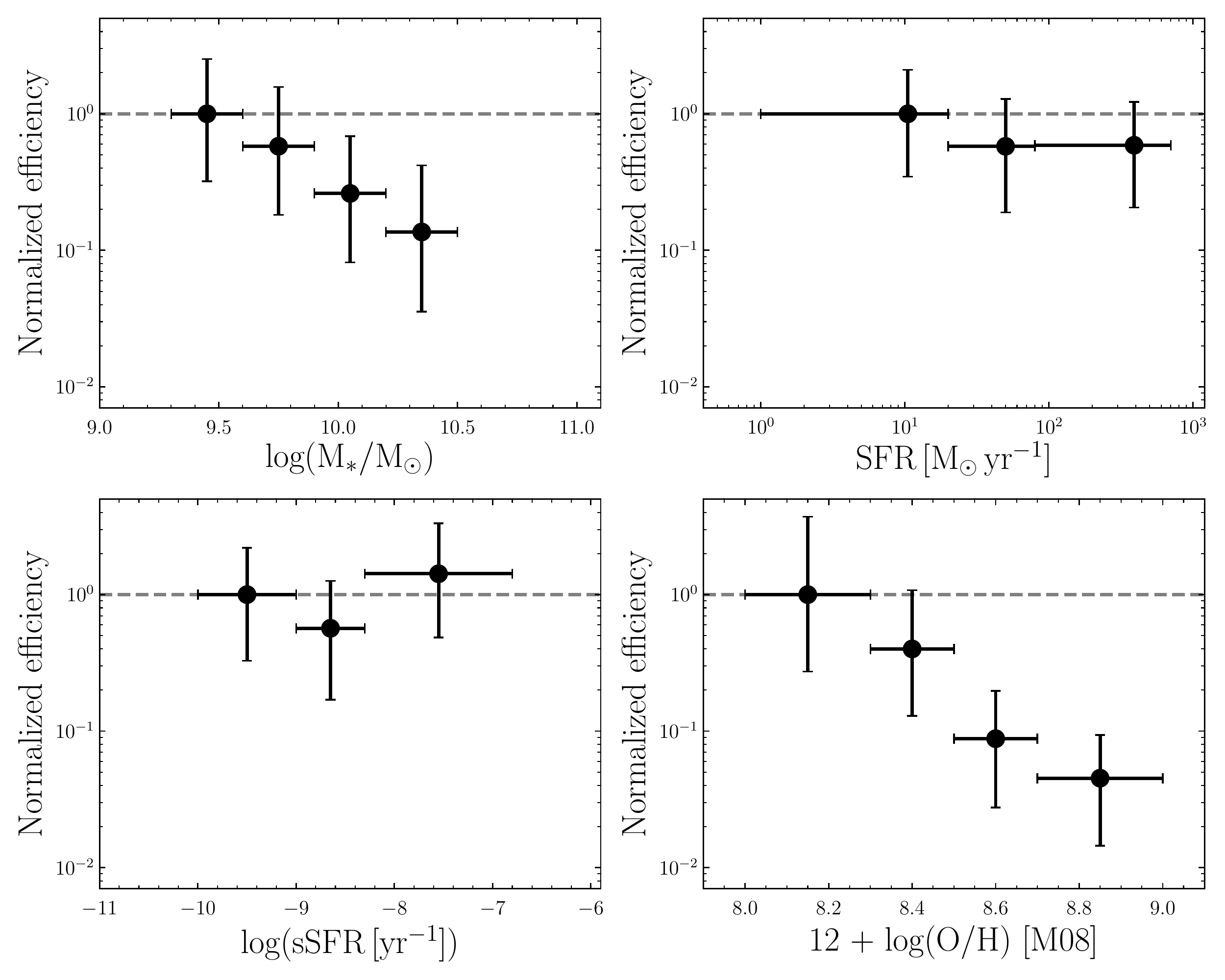}
\end{center}
\caption{
\small 
The normalised efficiency of LGRB hosts compared to the MOSDEF sample at \ozt\ as a function of stellar mass (top left panel), SFR (top right panel), sSFR (bottom left panel) 
and metallicity (bottom right panel).
The values are normalised to the first bin.
The horizontal grey dashed line indicates a value of 1 to guide the eye.}
\label{fig:b_vs_prop}
\end{figure}

\begin{figure}[!ht]
\centering
\includegraphics[width=0.5\textwidth]{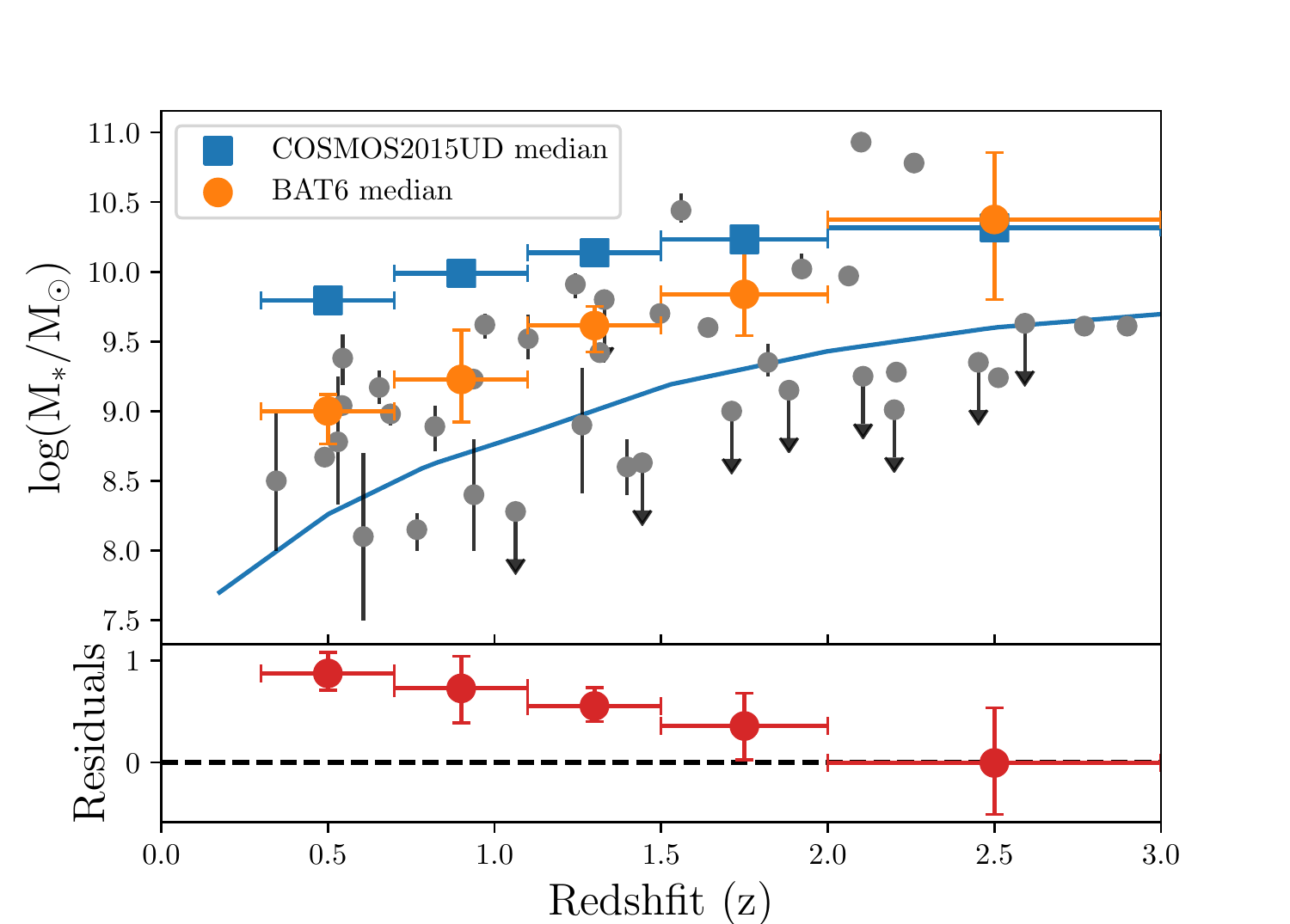}
\caption{
\small
Top panel: Stellar mass as a function of redshift.
The grey circles are the individual host galaxies of the BAT6 sample;
the orange circles represent the median stellar mass at each redshift bin for hosts above the COSMOS2015UD mass completeness.
The blue squares represent the median of the SFR-weighted stellar mass distribution of the COSMOS2015UD sample at each redshift bin.
The blue line is the mass completeness of the COSMOS2015UD sample.
Bottom panel: Residuals of the difference between the blue and orange points.
The errors are computed using Monte Carlo propagation and bootstrapping.
}
\label{fig:binned_stellar_mass}
\end{figure}

\begin{figure*}[!ht]
\begin{subfigure}[b]{0.48\textwidth}
\centering
\includegraphics[width=0.9\textwidth]{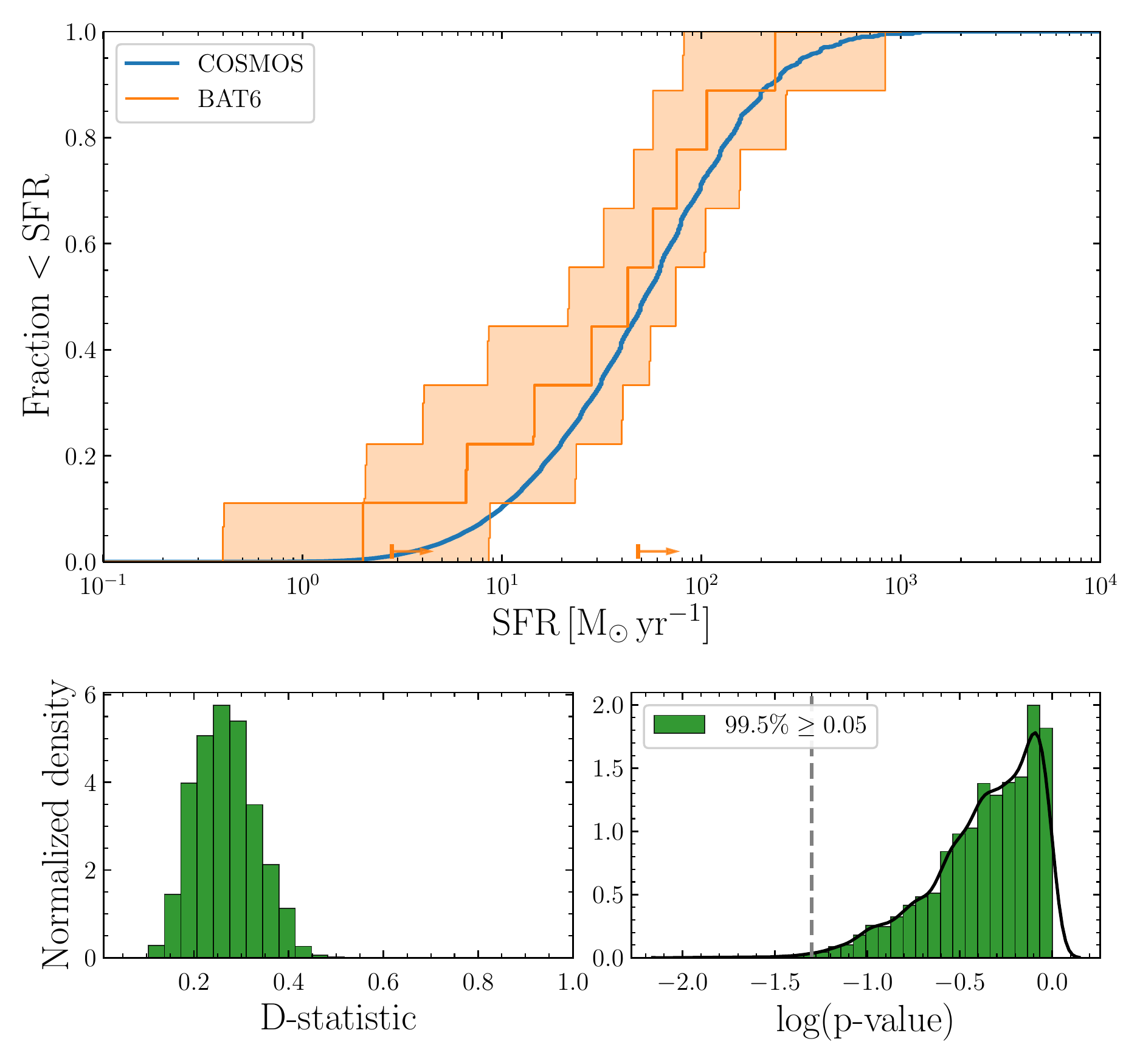}
\end{subfigure}
\hfill
\begin{subfigure}[b]{0.48\textwidth}
\centering
\includegraphics[width=0.9\textwidth]{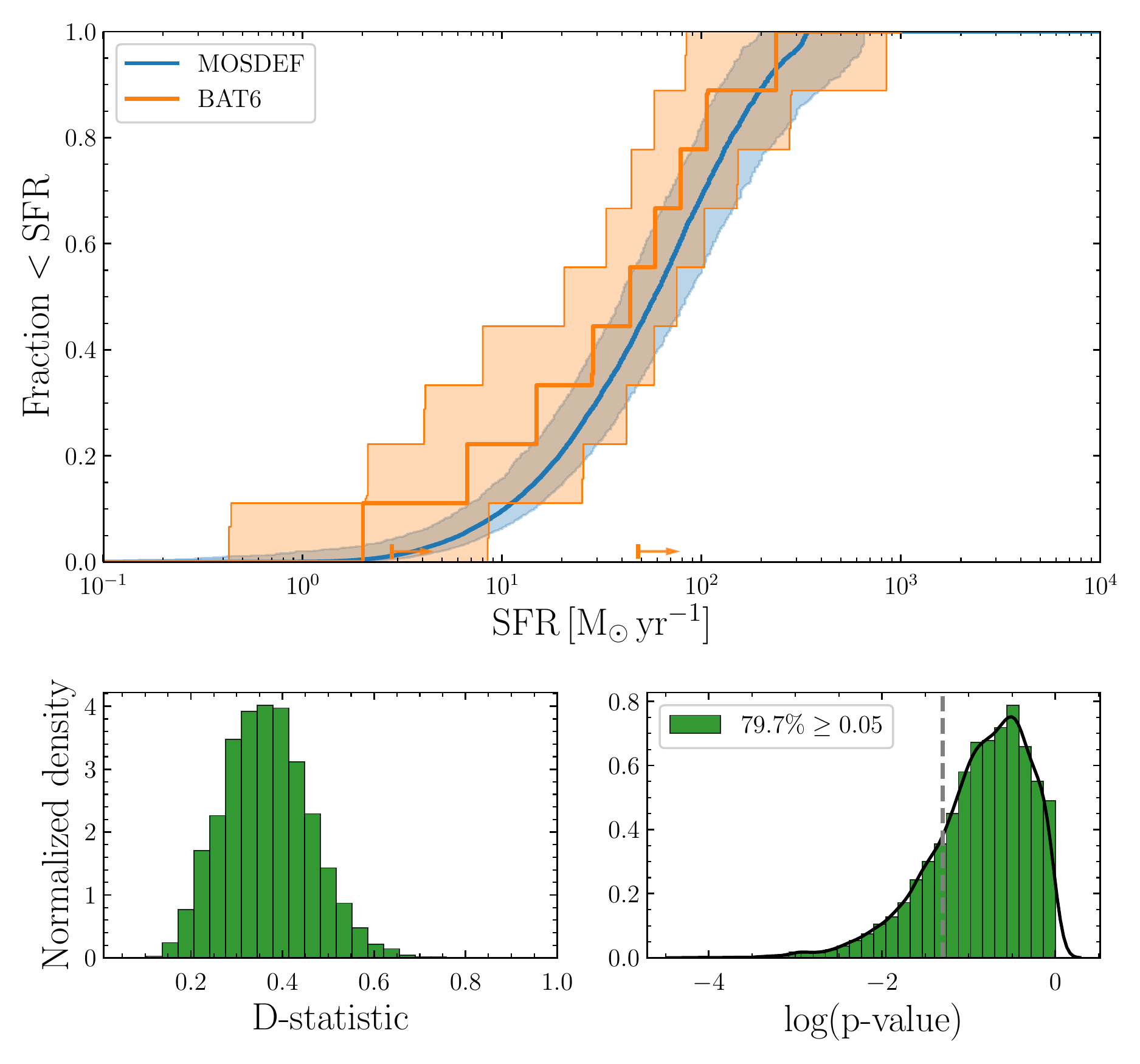}
\end{subfigure}
\caption{
\small
Top panels: Cumulative SFR distribution for the hosts of the BAT6 sample (orange) and the star-forming galaxies from the COSMOS2015 Ultra Deep catalogue (blue, left panel) and MOSDEF (blue, right panel) at \ozt.
The COSMOS2015 Ultra Deep and MOSDEF CDFs are weighted by SFR.
Limits are indicated by arrows at the bottom of the plot.
Bottom panels: See Fig.\,\ref{fig:mass_CDF}.
}
\label{fig:SFR_CDF}
\end{figure*}

In the right panels of Figure~\ref{fig:mass_CDF}, the same comparison is performed with the star-forming galaxies of the MOSDEF survey. In this case the SFR used is that determined from the dust-corrected \Ha\, luminosities.
We computed 10000 MC realisations of both the BAT6 and the MOSDEF sample with the assumptions described in Sect.~\ref{subsec:Bayes}, with the difference that each galaxy in the MOSDEF sample is weighted by the realisation of its SFR.
For each realisation, we compute the 2 sample K-S test which yields a distribution of p-values firmly excluding the possibility that LGRB hosts are drawn from the same stellar mass distribution as that of MOSDEF galaxies weighted by their SFR.

Another way to look at the discrepancy of the distributions and have some information on the behaviour of the LGRB efficiency as a function of the stellar mass is to use the method presented by \cite{Boissier2013}, and used also in \cite{Vergani2015}. 
In the present work, instead of using galaxy models, in Fig.~\ref{fig:b_vs_prop} we compare the LGRB host galaxies directly with the MOSDEF star-forming galaxies.
The efficiency here is defined as the fraction of LGRB hosts divided by the fraction of MOSDEF galaxies in a given stellar mass or metallicity bin. 
The results are normalised to the first bin value.
We apply this method also for the galaxy properties presented in the following sections (SFR, sSFR, and metallicity; see Fig.~\ref{fig:b_vs_prop}).

We also investigated the evolution of the median stellar mass with redshift for the BAT6 hosts compared to the SFR-weighted COSMOS2015UD sample, presented in Fig.~\ref{fig:binned_stellar_mass}.
The discrepancy between the BAT6 hosts and the SFR-weighted field galaxies is most notable at low redshift and decreases up to $z=3$ as is shown in the bottom panel, although the last redshift bin is to be taken cautiously due the low number of hosts within it.
Additionally, the stellar masses of the LGRB hosts in the last redshift bin ($2 < z < 3$) are derived using a different methodology (see~\ref{subsec:prop_M*}).
With these caveats in mind, this trend is consistent with the observations of \citealt{Perley2016} (see also \citealt{Hunt2014}).

\subsection{Star formation rate}\label{subsec:survey_compar_SFR}

\begin{figure*}[!ht]
\begin{subfigure}[b]{0.48\textwidth}
\centering
\includegraphics[width=0.9\textwidth]{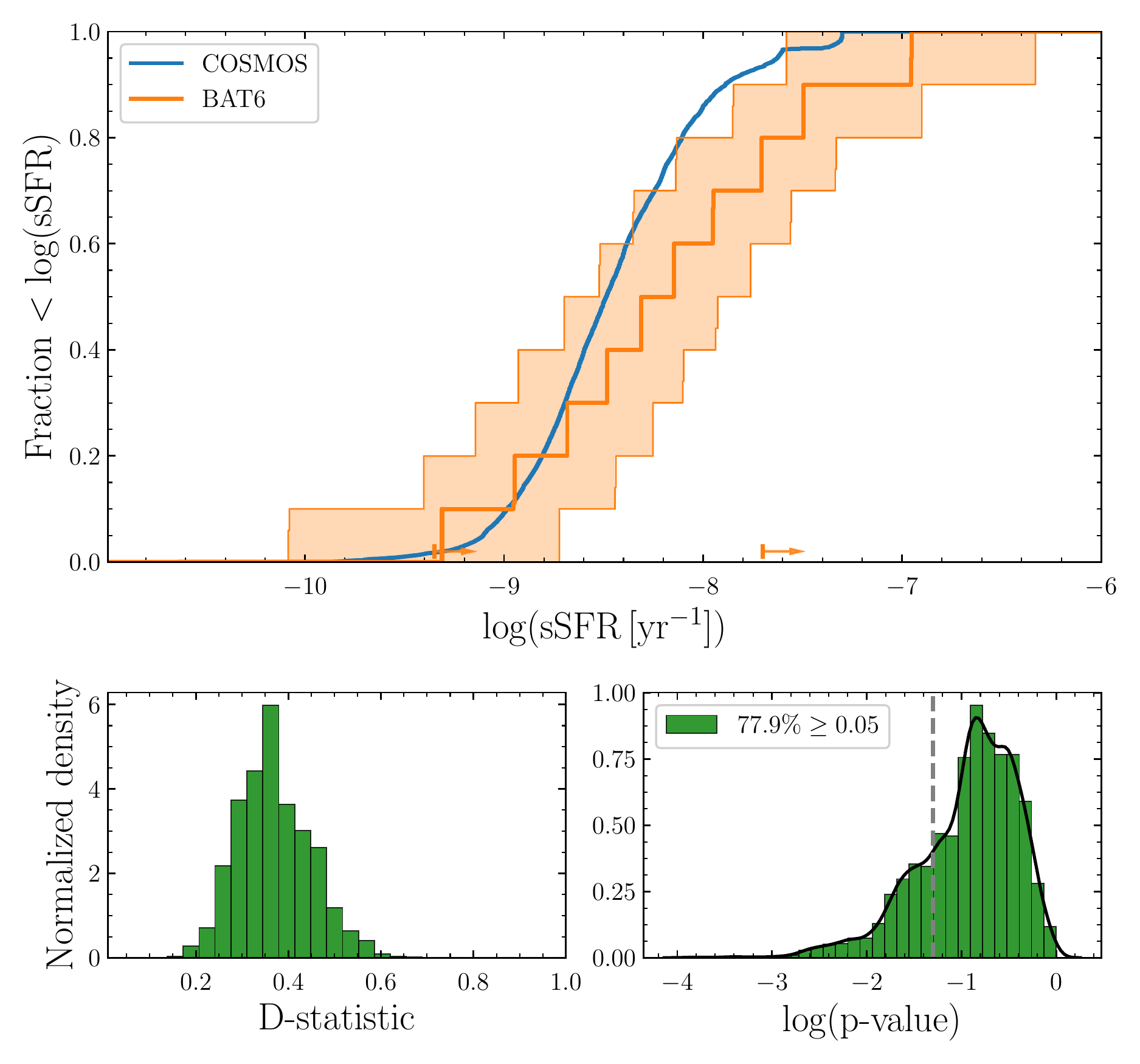}
\end{subfigure}
\hfill
\begin{subfigure}[b]{0.48\textwidth}
\centering
\includegraphics[width=0.9\textwidth]{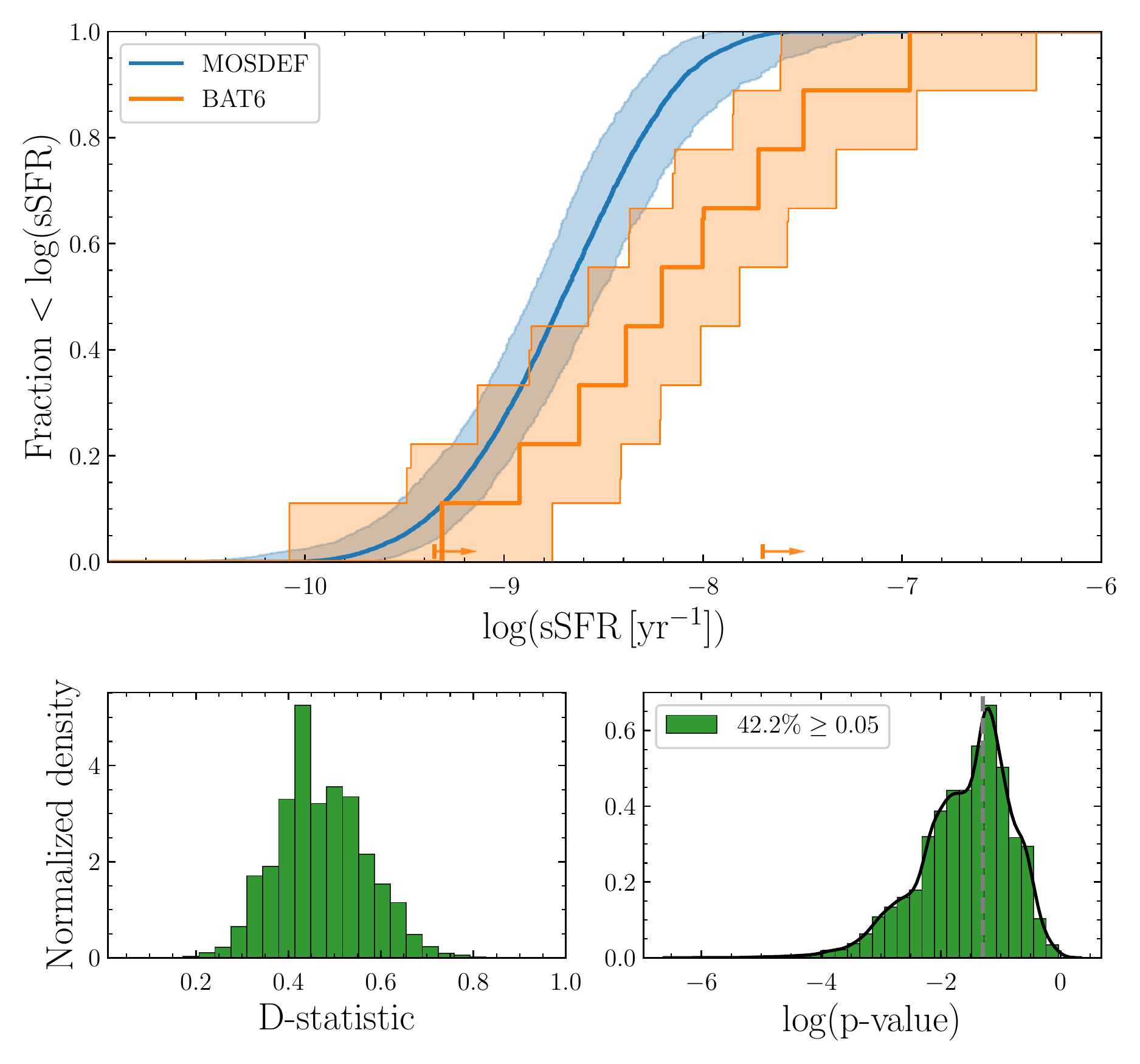}
\end{subfigure}
\caption{
\small
Top panels: Cumulative sSFR distribution for the hosts of the BAT6 sample (orange) and the star-forming galaxies from the COSMOS2015 Ultra Deep catalogue (blue, left panel) and MOSDEF (blue, right panel) at \ozt.
The COSMOS2015 Ultra Deep and MOSDEF CDFs are weighted by SFR.
Limits are indicated by arrows at the bottom of the plot.
Bottom panels: See Fig.~\ref{fig:mass_CDF}
}
\label{fig:sSFR_CDF}
\end{figure*}

The top panels of Figure \ref{fig:SFR_CDF} show the SFR cumulative distribution of hosts of the BAT6 sample compared to SFR-weighted 
distribution of star-forming field galaxies of COSMOS2015UD and MOSDEF at \ozt. 
As confirmed by the p-value distribution, there is good agreement between the two distributions.

\begin{figure}[!ht]
\begin{center}
\includegraphics[width=0.9\linewidth]{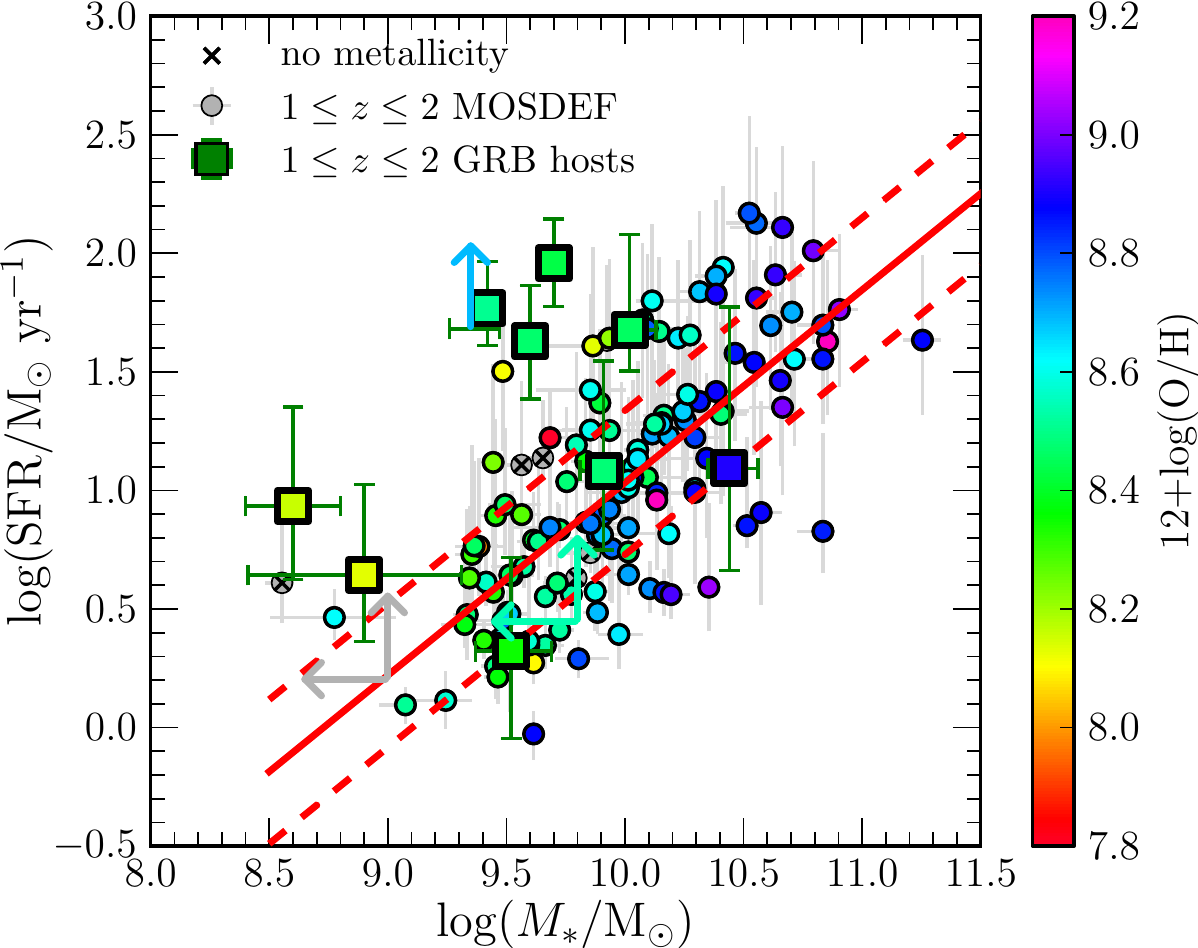}
\end{center}
\caption{
\small
SFR versus stellar mass plot for the BAT6 sample at $1<z<2$ (squares).
The circles are from the MOSDEF sample at \ozt. 
The red line is the best fit to the MOSDEF data and the dotted lines represent the intrinsic scatter, following the method of \citet{Shivaei2015}. 
The points are coloured by metallicity in the M08 calibrator.
Galaxies where no metallicity could be measured are coloured in grey with a cross.
}
\label{fig:SFMS}
\end{figure}

\begin{figure}
\begin{center}
\includegraphics[width=0.8\linewidth]{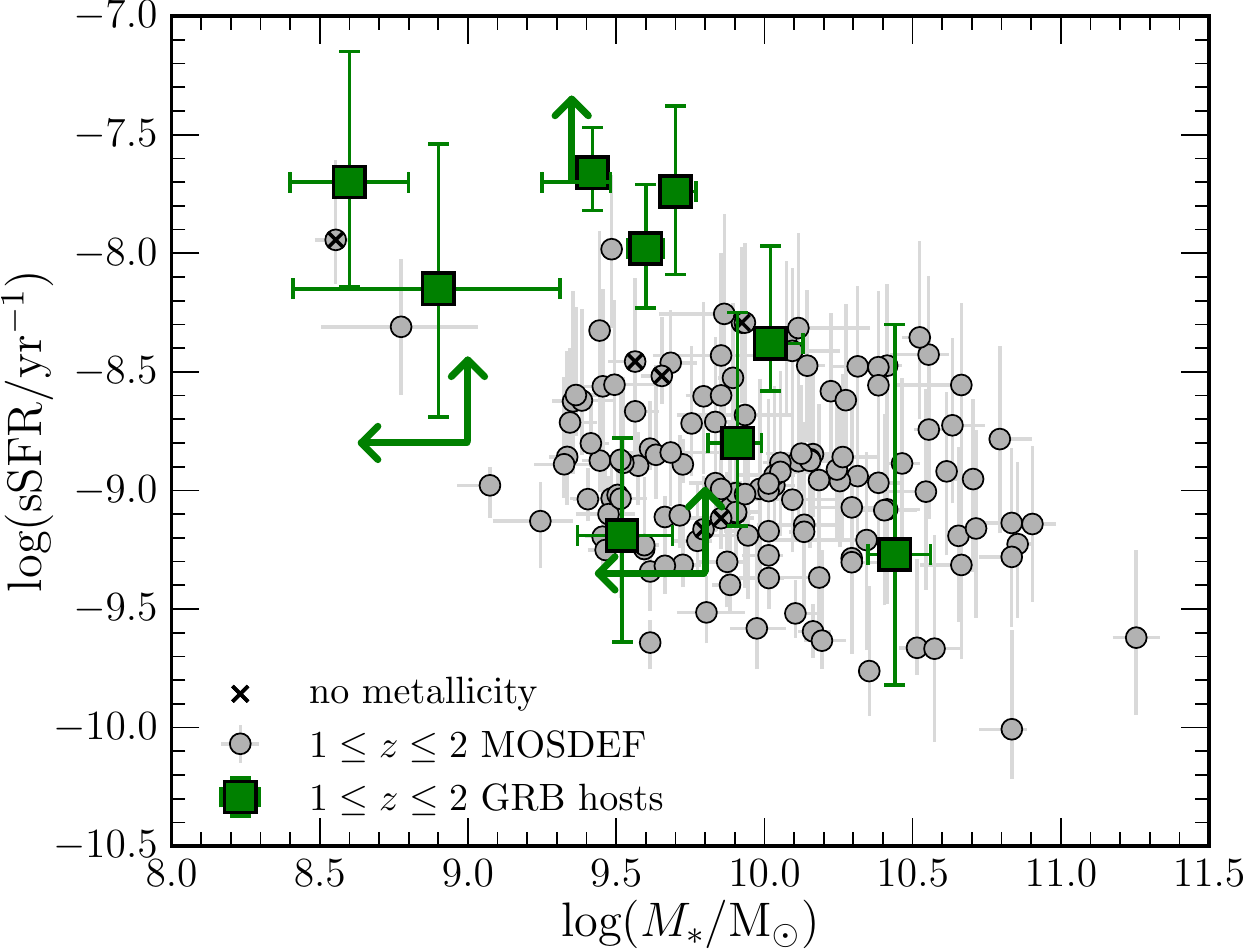}
\end{center}
\caption{
\small
sSFR versus stellar mass plot for the BAT6 sample at \ozt. The grey circles are from the MOSDEF sample at \ozt.}
\label{fig:sSFMS}
\end{figure}

The top panels of Figure \ref{fig:sSFR_CDF} show the specific SFR (sSFR, defined as SFR/$\mathrm{M_*}$) cumulative distribution of hosts of the BAT6 sample compared to the SFR-weighted distribution of star-forming field galaxies of COSMOS2015UD and MOSDEF at \ozt. 
The p-value distribution in the bottom panels indicates that in the majority of cases we cannot exclude the null hypothesis that the two samples are drawn from the same distribution for the COSMOS2015UD sample, while for the MOSDEF sample, it is less definitive since the p-value distribution peaks around 0.05.
In $\sim$ 40\% of cases, we cannot discard the null hypothesis at the 95\% confidence.

We note that we could not determine the SFR for three host galaxies. Nonetheless their stellar masses were lower than the stellar mass completeness of the surveys. Therefore,  when comparing with surveys our sample is still complete.

In Figs.~\ref{fig:SFMS} and \ref{fig:sSFMS} we plot the BAT6 host galaxies and the MOSDEF star-forming galaxies in the SFR, sSFR vs stellar mass plane, respectively.
We fit the SFR vs stellar mass relation (so-called \textit{Main Sequence}, e.g. \citealt{Whitaker2012}) for the MOSDEF sample of star-forming galaxies at \ozt\ following \citet{Shivaei2015}.
We derived the fraction of galaxies above the 1-sigma intrinsic scatter (see \citealt{Japelj2016a}) of the relation within the MOSDEF sample to be 27$\pm$5\%.
Excluding the 3 hosts falling in the low mass region, sparsely populated by the MOSDEF sample, the fraction of LGRB host galaxies showing such an enhancement of SFR, with respect to the MOSDEF \ozt\ relation, is 66$\pm$22\%.

\subsection{Metallicity}\label{subsec:survey_compar_logOH}

\begin{figure}[!ht]
\begin{center}
\resizebox{\hsize}{!}{\includegraphics{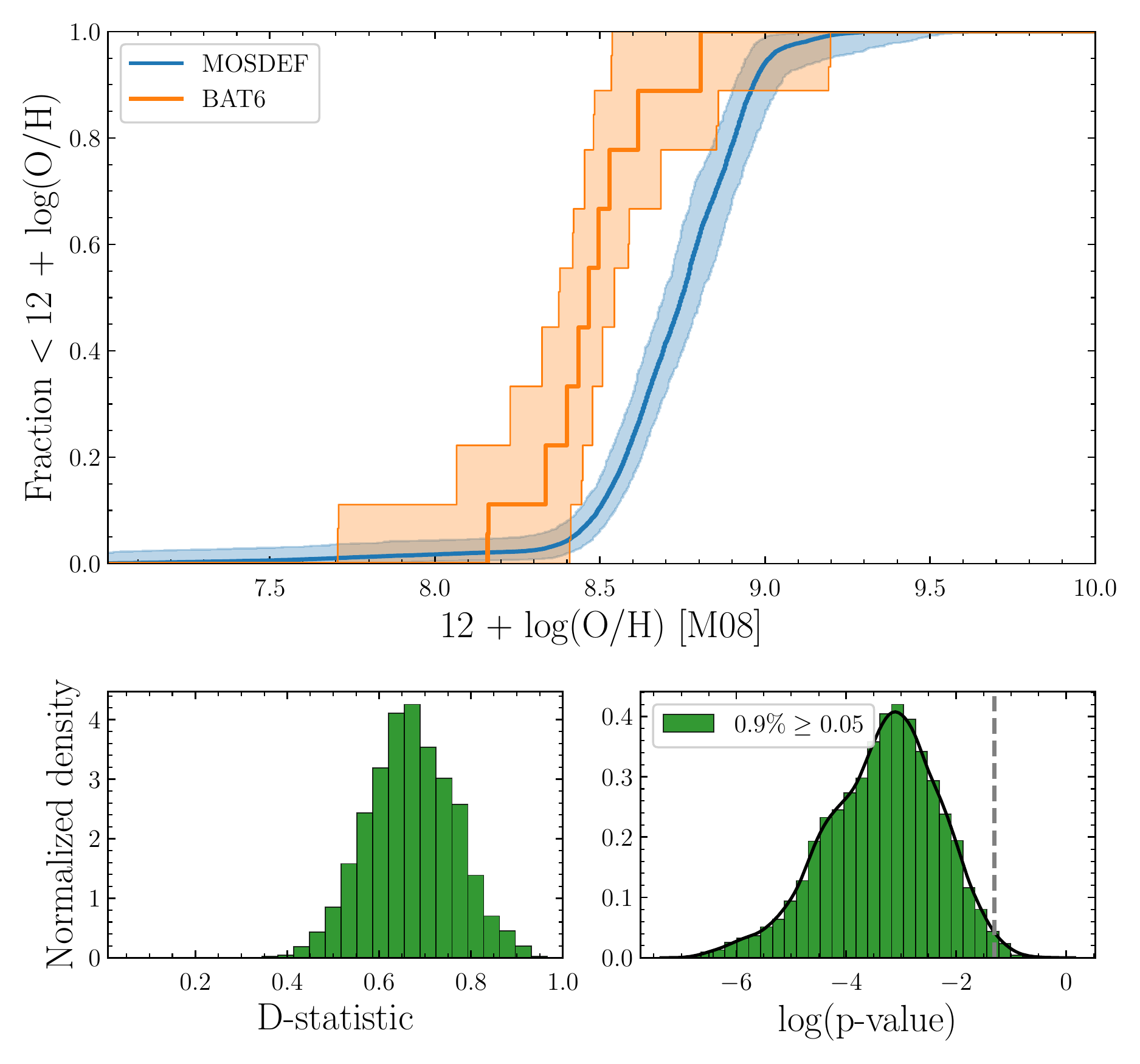}}
\end{center}
\caption{
\small
Top panel: Cumulative metallicity distribution for the hosts of the BAT6 sample (orange) and the star-forming galaxies from the MOSDEF sample (blue) at \ozt.
The MOSDEF CDF is weighted by SFR.
Bottom panels: See Fig.\,\ref{fig:mass_CDF}.
}
\label{fig:logOH_CDF_MOSDEF}
\end{figure}

The MOSDEF survey allows us also to perform the comparison of the metallicity distribution, within the same redshift range and using the same calibrator (M08). 
Fig. \ref{fig:logOH_CDF_MOSDEF} shows the cumulative distribution of the metallicity of hosts of the BAT6 sample compared to the SFR-weighted distribution of star-forming galaxies of the MOSDEF at \ozt.
The distribution of p-values in the bottom right panel indicates we can reject the hypothesis that the MOSDEF star-forming galaxy sample weighted by SFR and the BAT6 sample are drawn from the same distribution at the 95\% confidence level.

We note that we could not determine the metallicity for four host galaxies. Nonetheless their stellar masses were lower than the stellar mass completeness of the MOSDEF sample. Therefore, our sample is still complete with respect to the comparison with the MOSDEF galaxies.

\begin{figure}
\begin{center}
\includegraphics[width=0.8\linewidth]{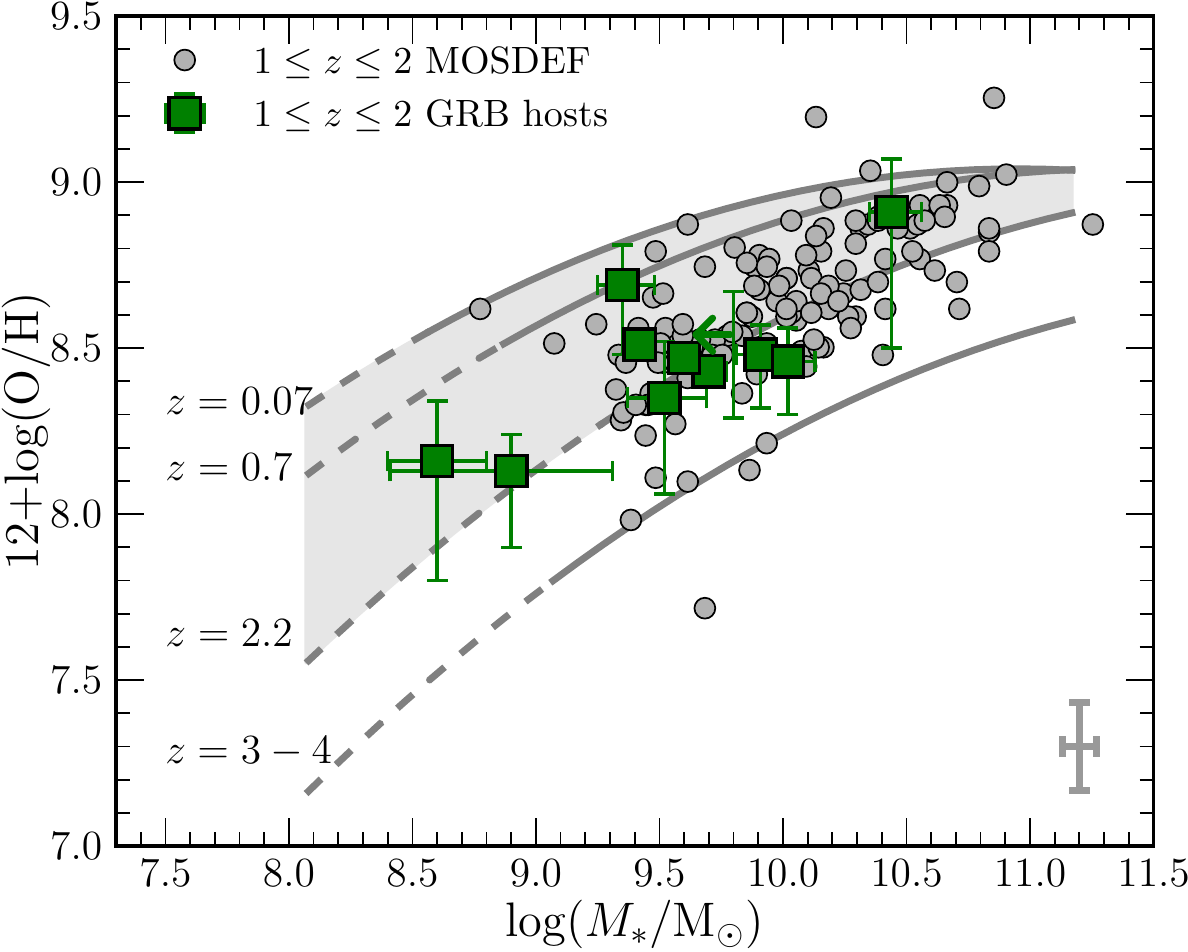}
\end{center}
\caption{
\small 
MZR in the M08 calibrator for the BAT6 sample at \ozt\ (squares). 
The grey points are from the MOSDEF sample at \ozt, with their average uncertainty shown on the bottom right.
The curves represent the MZR relation of \citet{Mannucci2009} from $z=0.07$ to $z\sim 3.5$, with the extrapolation below the mass completeness indicated in dashed.
}
\label{fig:MZR}
\end{figure}

Figure~\ref{fig:MZR} shows the mass-metallicity relation (MZR) for the BAT6 hosts and the MOSDEF sample, using the M08 calibrator.
We see that the LGRB hosts are consistent with the star-forming field galaxies at low mass and low metallicity but there is a clear dearth of high mass and high metallicity LGRB host galaxies\footnote{In \cite{Vergani2017} the authors also present the MZR based on the same BAT6 sample but the stellar masses are revised in this work (see Sect.~\ref{subsec:prop_M*}).}.
Indeed there is only one host (which has very large errors) above \logOH\ $\sim$ 8.7, whereas the area of stellar masses above $\sim$ 10$^{10}$\,\Msun\,and \logOH\ $\sim$ 8.6 is well populated by the star-forming galaxies of MOSDEF.

\begin{figure}
\begin{center}
\includegraphics[width=0.8\linewidth]{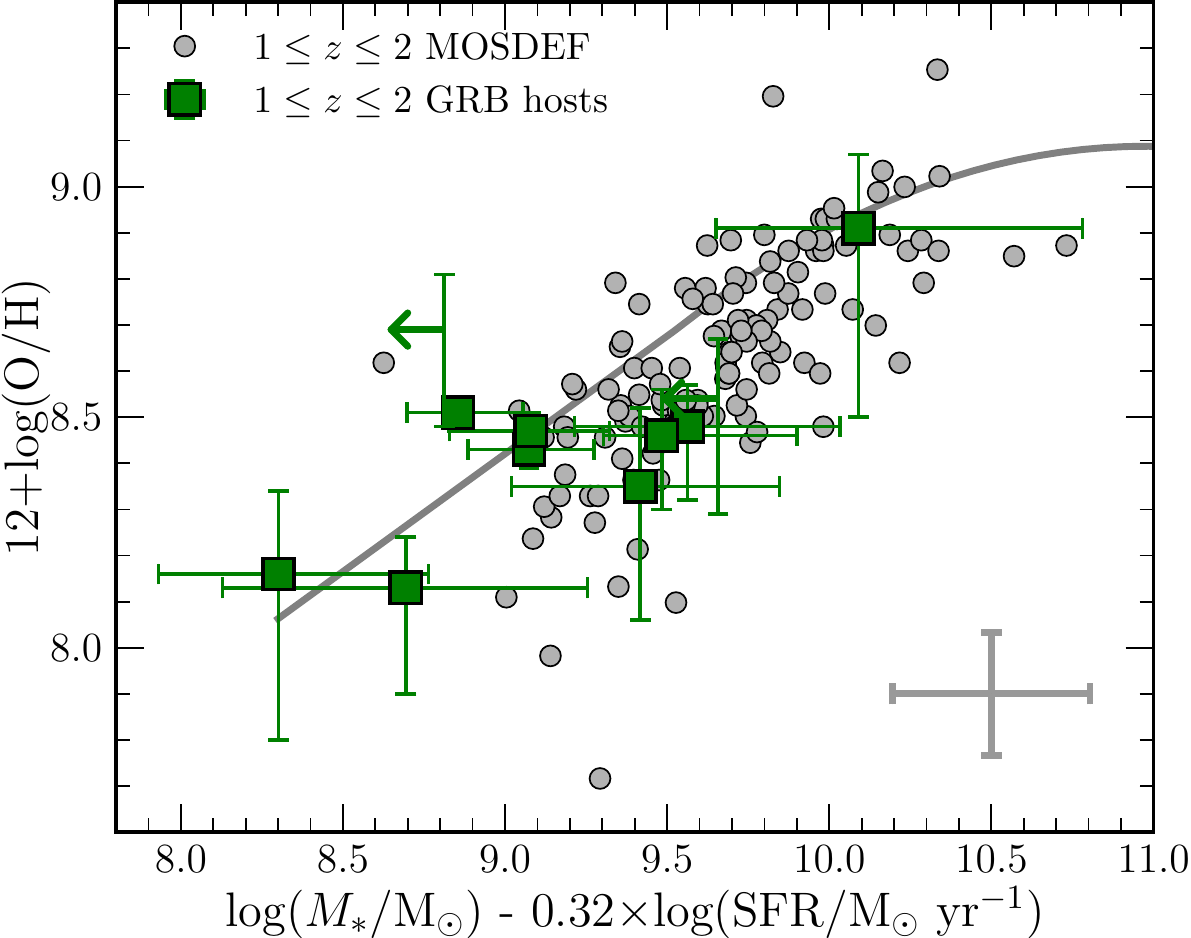}
\end{center}
\caption{
\small 
FMR for the BAT6 sample at \ozt\ (squares). The grey points are from the MOSDEF sample at \ozt, with their average uncertainty shown on the bottom right. The grey line is the FMR from \citet{Mannucci2011}.}
\label{fig:FMR}
\end{figure}

We also computed the Fundamental Metallicity Relation (FMR) as defined by \citet{Mannucci2011}, represented in Fig.~\ref{fig:FMR}.
This relation is supposed to be redshift independent. Nonetheless, as \cite{Sanders2015, Sanders2018} find a redshift dependence of the FMR built with the MOSDEF sample, we prefer to plot here only the BAT6 hosts at \ozt, omitting hosts at $z<1$.

In \cite{Vergani2017} the authors noted a discrepancy between the region occupied by the LGRB hosts and the FMR (explained by a metallicity threshold for LGRB production).
Here, it appears the LGRB hosts occupy mostly the low $\mu$\footnote{Where $\mu = $ \logM\ - 0.32\,$\mathrm{\log(SFR/M_{\odot}\,yr^{-1})}$} area (whereas roughly half of the MOSDEF sample lies at $\mu > 9.7$), and, in this region, they are consistent with the MOSDEF points. 
However, at those $\mu$ values, both the MOSDEF sample and the LGRB hosts seem to have lower metallicities with respect to the FMR predictions. 
A complete analysis of this discrepancy is beyond the scope of this paper. 
Here we can point out that this could be due to an underestimation of the FMR slope at low $\mu$, or to an evolution of the relation in redshift (as found by \citealt{Sanders2018}), as we are comparing galaxies at \ozt\, (our LGRB and MOSDEF samples) with the FMR built mainly with low-redshift galaxies. 
Indeed, different works showed that evolving physical conditions of ionised gas in HII regions may lead to evolution in the relationships between emission-line ratios and metallicity (e.g. \,\citealt{Steidel2014,Shapley2015,Sanders2016}). 
This is not an issue when comparing the metallicities of the BAT6 sample and the MOSDEF one as we selected the same redshift range \ozt, unless the physical conditions in LGRB hosts are significantly different from those in typical SF MOSDEF galaxies.  
 
\section{Discussion}\label{sec:discussion}

The analysis presented in the previous sections clearly shows that the stellar mass and metallicity CDFs of the LGRB hosts do not follow those of typical star-forming galaxies weighted by SFR.
This implies that, due to some factors affecting the LGRB production efficiency, at \ozt\ the LGRB rate cannot be used to directly trace star formation.
As found in previous work \citep{Vergani2015,Perley2016b,Japelj2016,Vergani2017}, it seems that metallicity is the main factor involved: LGRBs explode preferentially in sub-solar metallicity environments. Indeed, as we will discuss in more detail later in this section, in the commonly used LGRB collapsar progenitor model \citep{Woosley1993} a dependence of the LGRB production on metallicity is expected. In this context, the discrepancies in the stellar mass distribution are a direct consequence of the relation between stellar mass and metallicity (lower metallicities correspond to lower stellar masses).

\begin{figure}
\begin{center}
\includegraphics[width=0.8\linewidth]{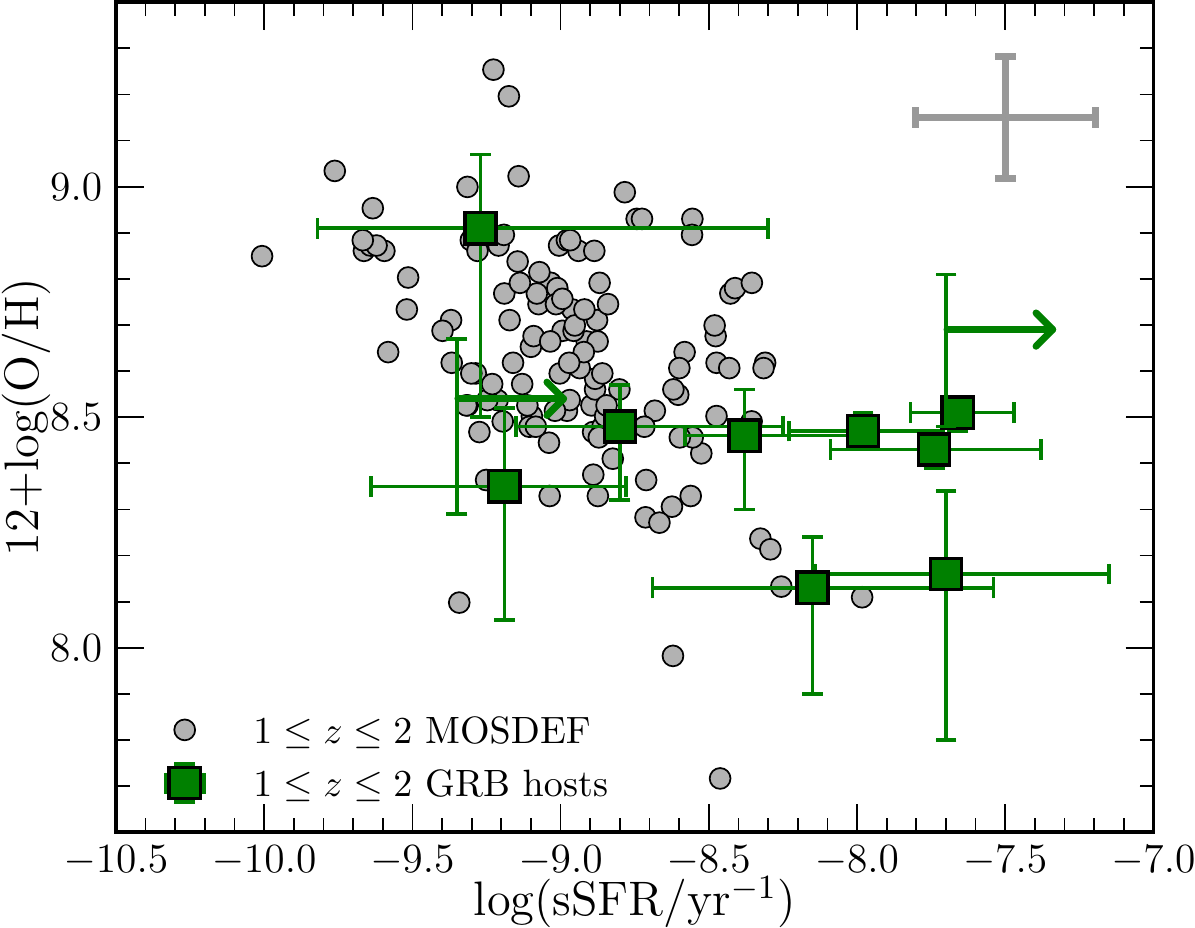}
\end{center}
\caption{
\small 
Metallicity-sSFR relation for the BAT6 sample at \ozt\ (squares). The grey points are from the MOSDEF sample at \ozt, with their average uncertainty shown on the upper right.}
\label{fig:logOH_vs_sSFR}
\end{figure}

\begin{figure*}[!ht]
\begin{subfigure}[b]{0.48\textwidth}
\centering
\includegraphics[width=0.9\textwidth]{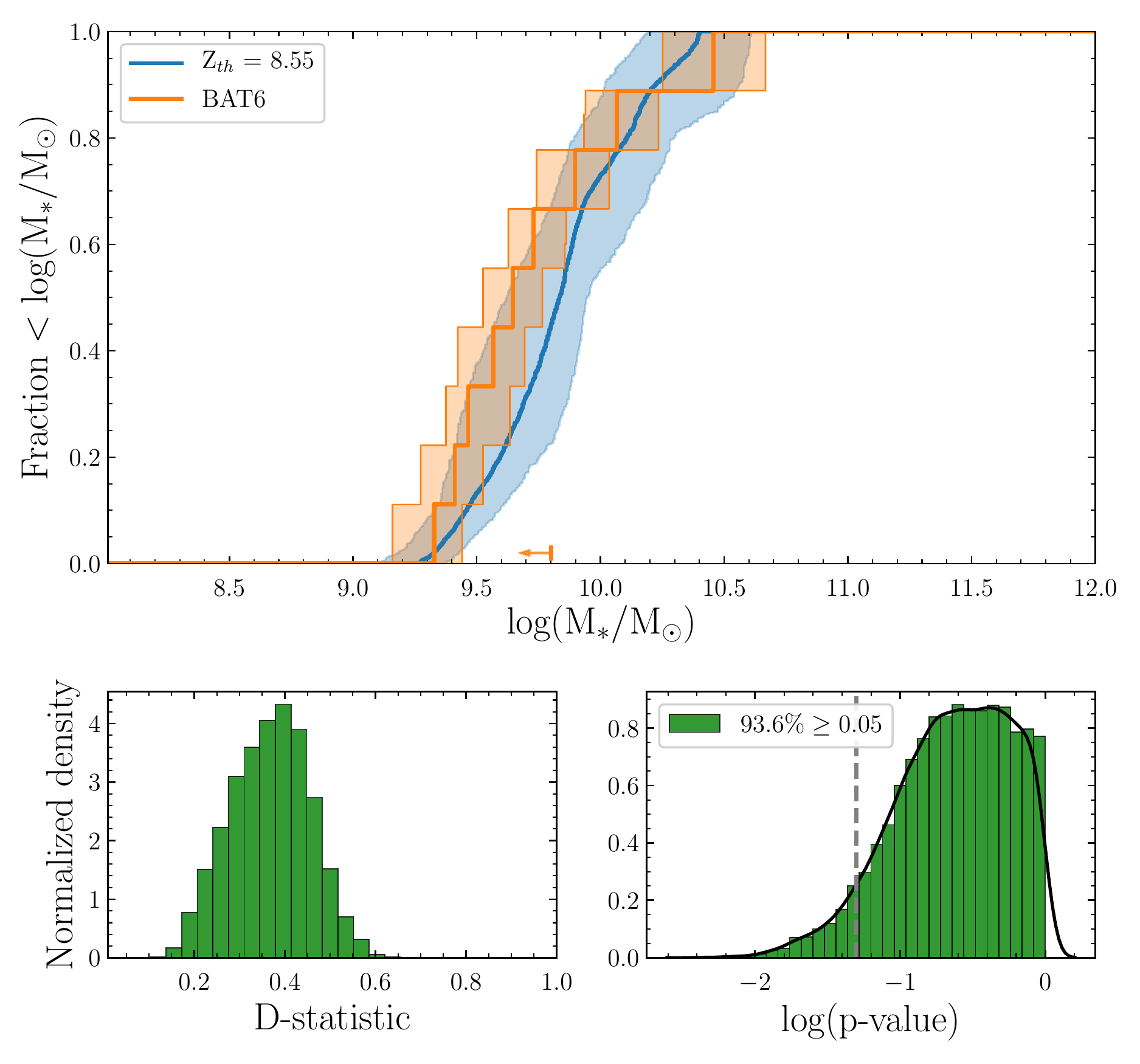}
\end{subfigure}
\hfill
\begin{subfigure}[b]{0.48\textwidth}
\centering
\includegraphics[width=0.9\textwidth]{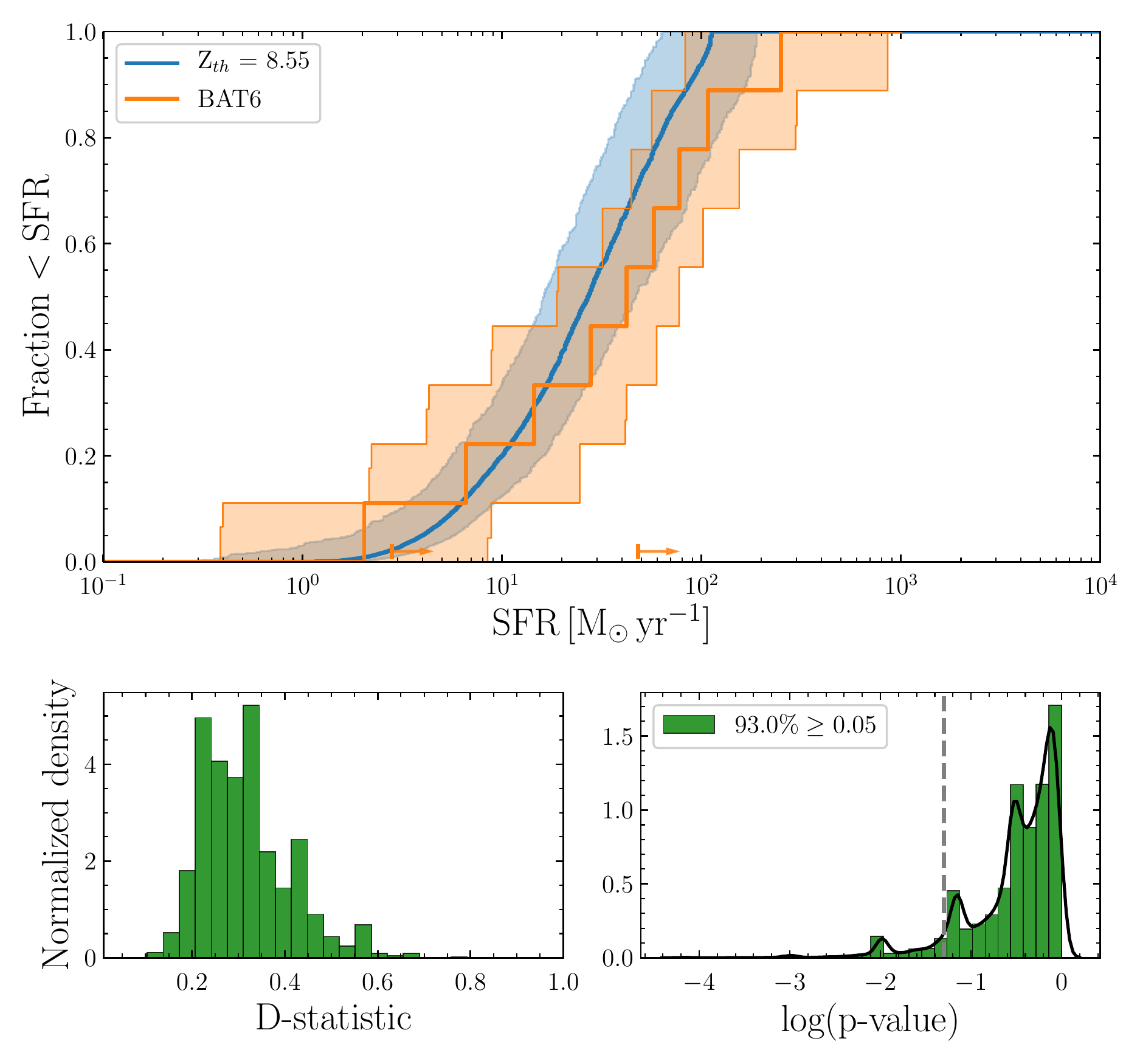}
\end{subfigure}
\begin{subfigure}[b]{0.48\textwidth}
\centering
\includegraphics[width=0.9\textwidth]{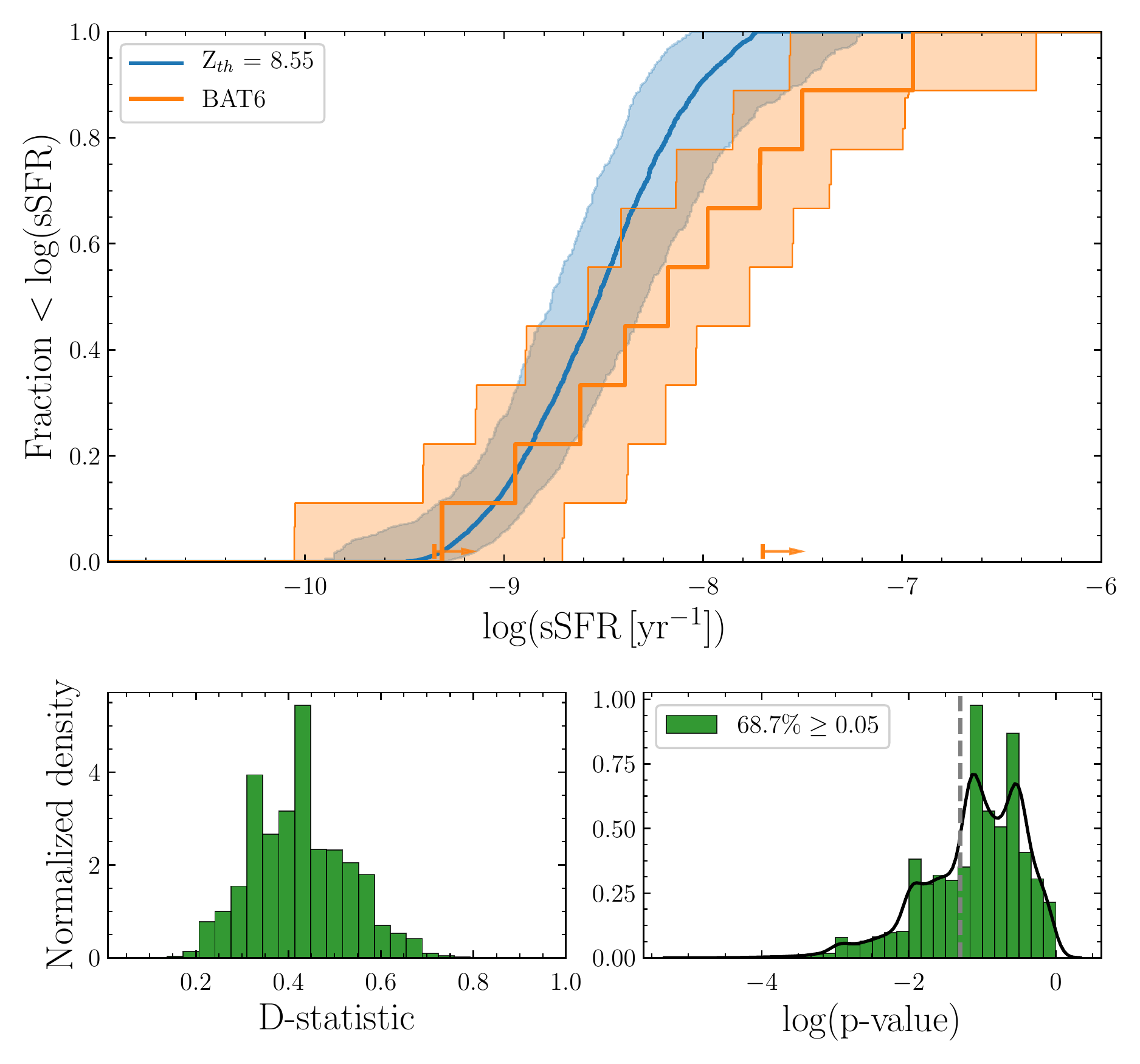}
\end{subfigure}
\hfill
\begin{subfigure}[b]{0.48\textwidth}
\centering
\includegraphics[width=0.9\textwidth]{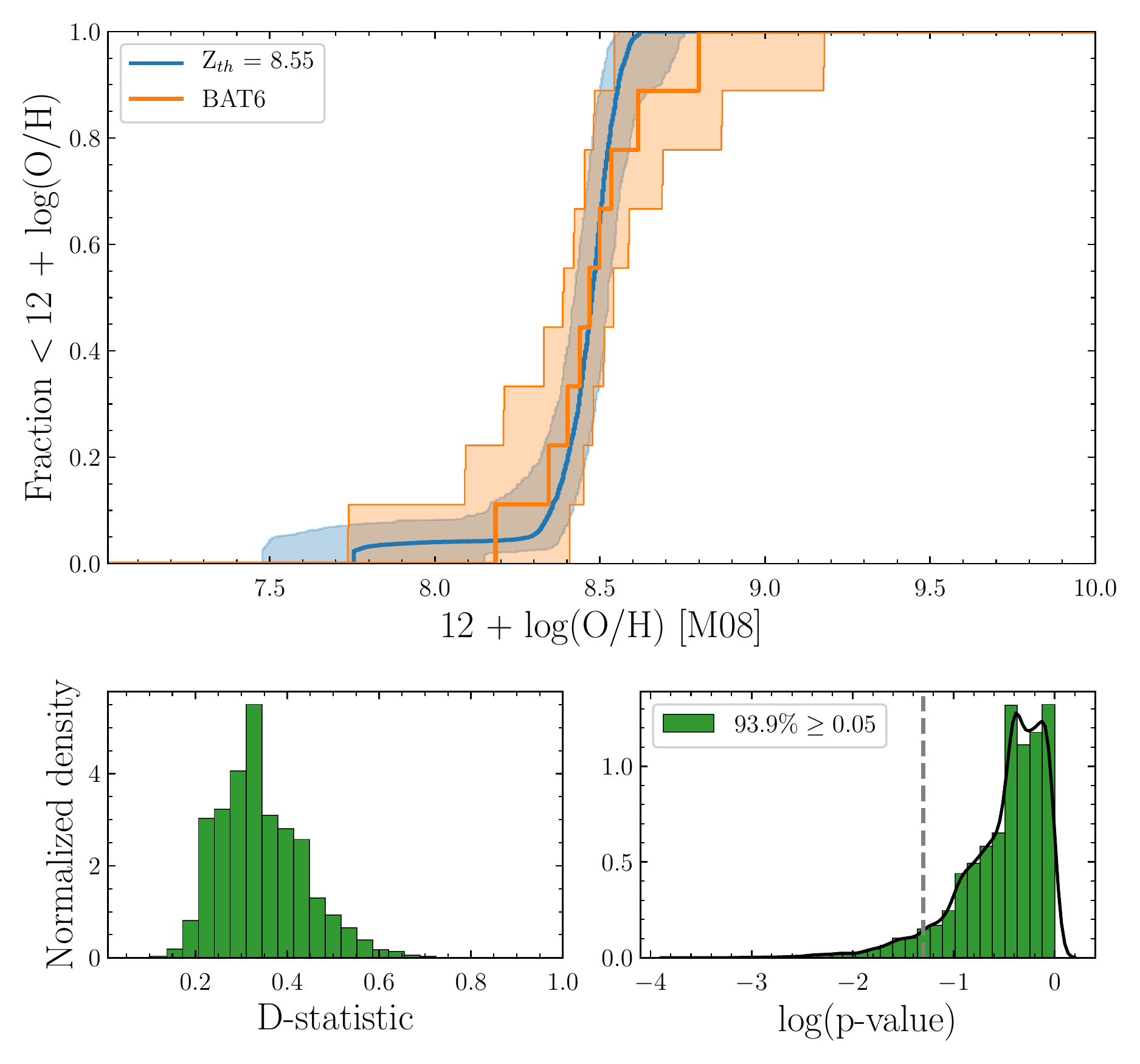}
\end{subfigure}
\caption{
\small
The result of the same analysis as presented in Section~\ref{sec:compar_samples} except using a metallicity cut on the MOSDEF sample of \logOH=8.55.
The CDFs match more closely, and the 2 sample K-S tests suggest we can not discard the null hypothesis in the majority of realisations.
}
\label{fig:all_CDF_MOSDEF_cut}
\end{figure*}

Nonetheless, in our analysis there seems to be evidence also for an enhancement of sSFR among LGRB host galaxies compared to star-forming galaxies found in galaxy surveys.
In the literature there are indications that starburst galaxies are generally characterised by lower metallicity than non-starburst ones (e.g. \citealt{Sanders2018}). 
It is therefore necessary to investigate which is the real driving factor affecting the LGRB efficiency, i.e. if it is the preference for galaxies with enhanced SFR that has as a consequence the preference for sub-solar metallicities, or the opposite.

Fig.~\ref{fig:logOH_vs_sSFR} shows that MOSDEF host galaxies with high sSFR have on average lower metallicity than those with lower sSFR values.
Nonetheless, within the sSFR range covered by the MOSDEF galaxies considered in this work, for a fixed sSFR the fraction of MOSDEF star-forming galaxies having metallicities larger than \logOH $\sim8.5$ is much higher than that of LGRB hosts. Stronger evidence that a possible preference for enhanced SFR would not be the only factor at play comes from the lack of LGRB host galaxies in the high stellar mass - high SFR region of Fig.~\ref{fig:SFMS}, compared to MOSDEF galaxies. Indeed, if enhanced SFR is the driving factor, we should find LGRB host galaxies with enhanced SFR also at stellar masses larger than $\sim$ 10$^{10}$\,\Msun.

All the results point towards metallicity as the main driving factor.
In order to further test this hypothesis, we apply a step-function metallicity cut on the MOSDEF sample and perform the comparison with our LGRB hosts again.
We impose different metallicity thresholds.
As the metallicity threshold value decreases the BAT6 and MOSDEF CDFs become more and more consistent until the majority of the p-values indicate we cannot confidently discard the null hypothesis that LGRB hosts and MOSDEF star-forming galaxies are drawn form the same population.
Using a metallicity cut of \logOH = 8.55, the SFR-weighted CDFs of MOSDEF come into agreement with the ones of the BAT6 sample, as is shown in Figure~\ref{fig:all_CDF_MOSDEF_cut}.
The two-sample K-S test results in a distribution of p-values consistent with the null hypothesis that the two samples are drawn from the same underlying distribution in the majority of MC realisations.
This implies that the discrepancies observed for the stellar mass and metallicity CDFs can be explained by a simple threshold on the metallicity, without the need for a contribution from a preference for starburst galaxies. This also naturally explains the trend observed in Fig.~\ref{fig:binned_stellar_mass}. Following the redshift evolution of the MZR, as the redshift increases, to a given stellar mass corresponds a lower metallicity. The metallicity threshold is therefore fulfilled by galaxies more and more massive. This explains the evolution of the median stellar mass of the LGRB host galaxies reaching the agreement with that of SF galaxies at $z\sim3$ (see also Section 5).

\begin{figure*}[!ht]
\begin{subfigure}[b]{0.48\textwidth}
\centering
\includegraphics[width=0.9\textwidth]{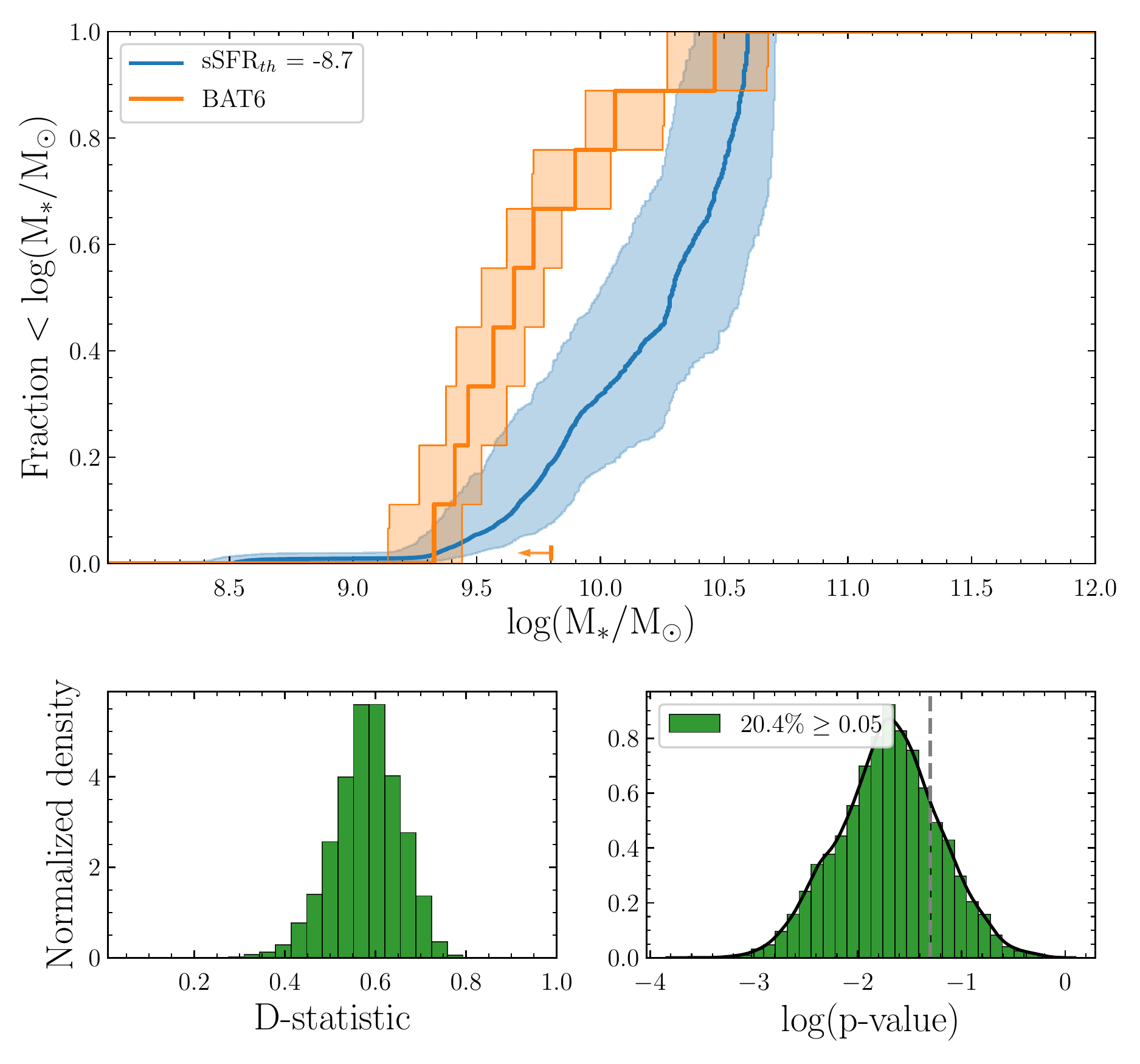}
\end{subfigure}
\hfill
\begin{subfigure}[b]{0.48\textwidth}
\centering
\includegraphics[width=0.9\textwidth]{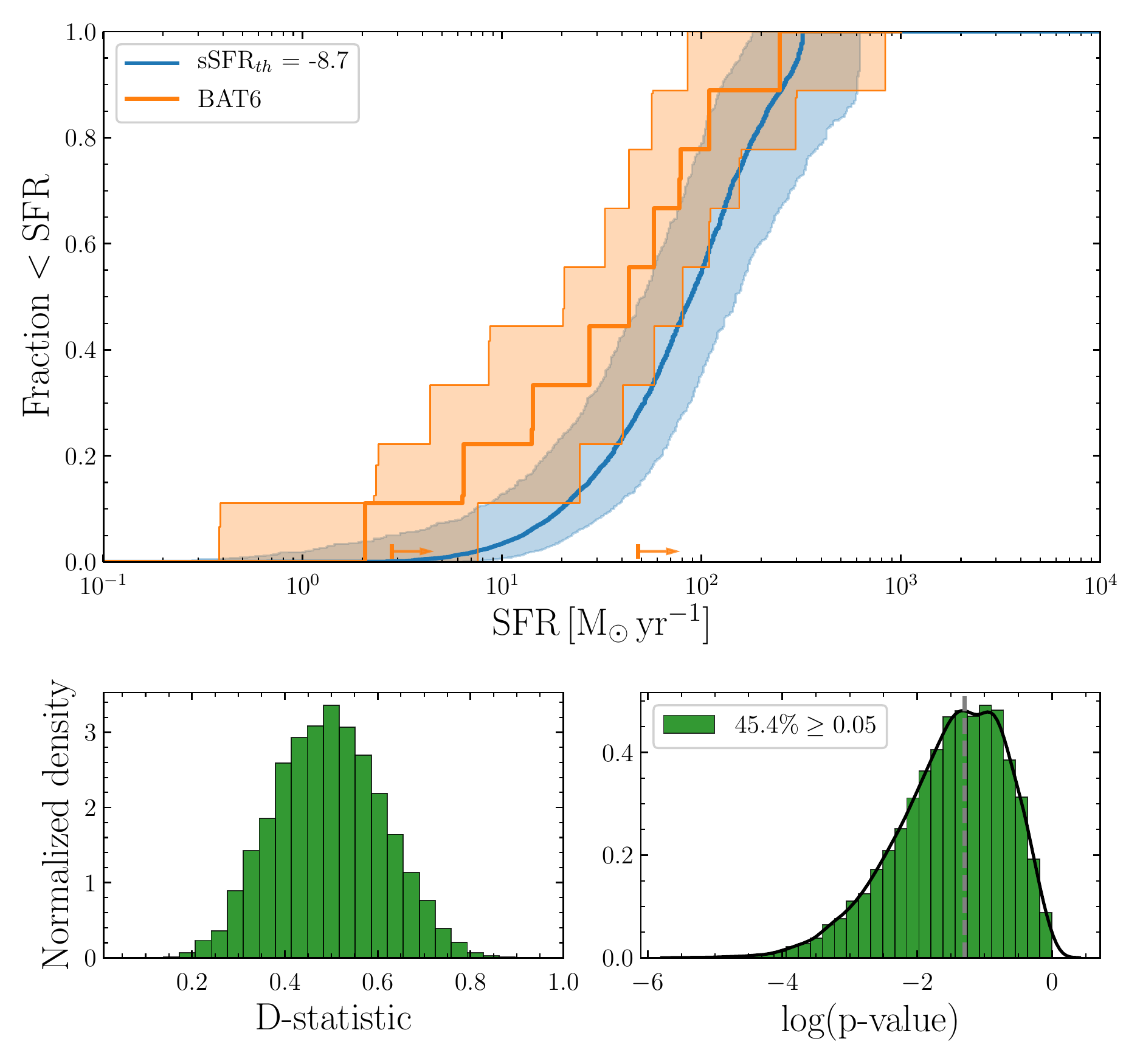}
\end{subfigure}
\begin{subfigure}[b]{0.48\textwidth}
\centering
\includegraphics[width=0.9\textwidth]{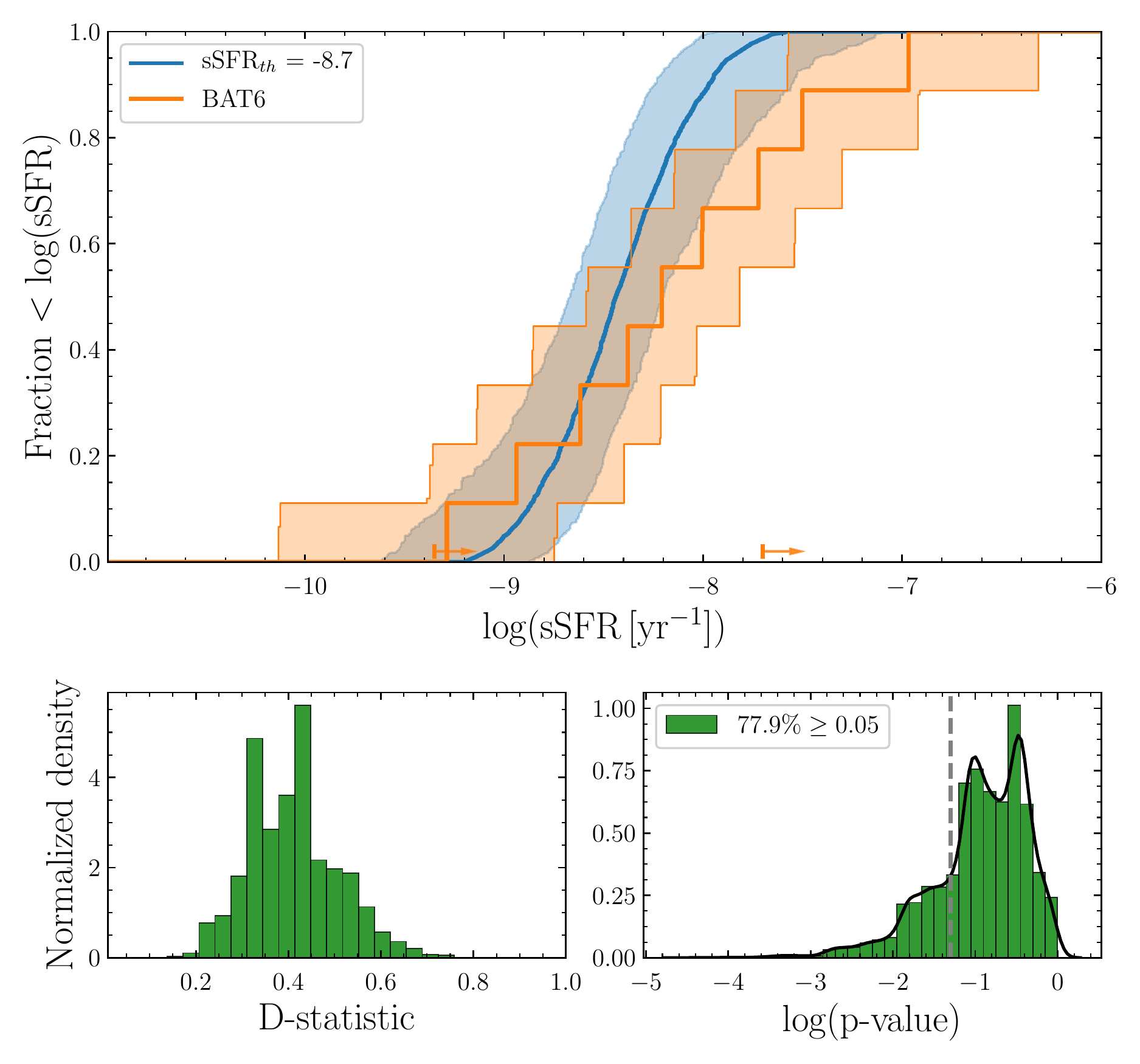}
\end{subfigure}
\hfill
\begin{subfigure}[b]{0.48\textwidth}
\centering
\includegraphics[width=0.9\textwidth]{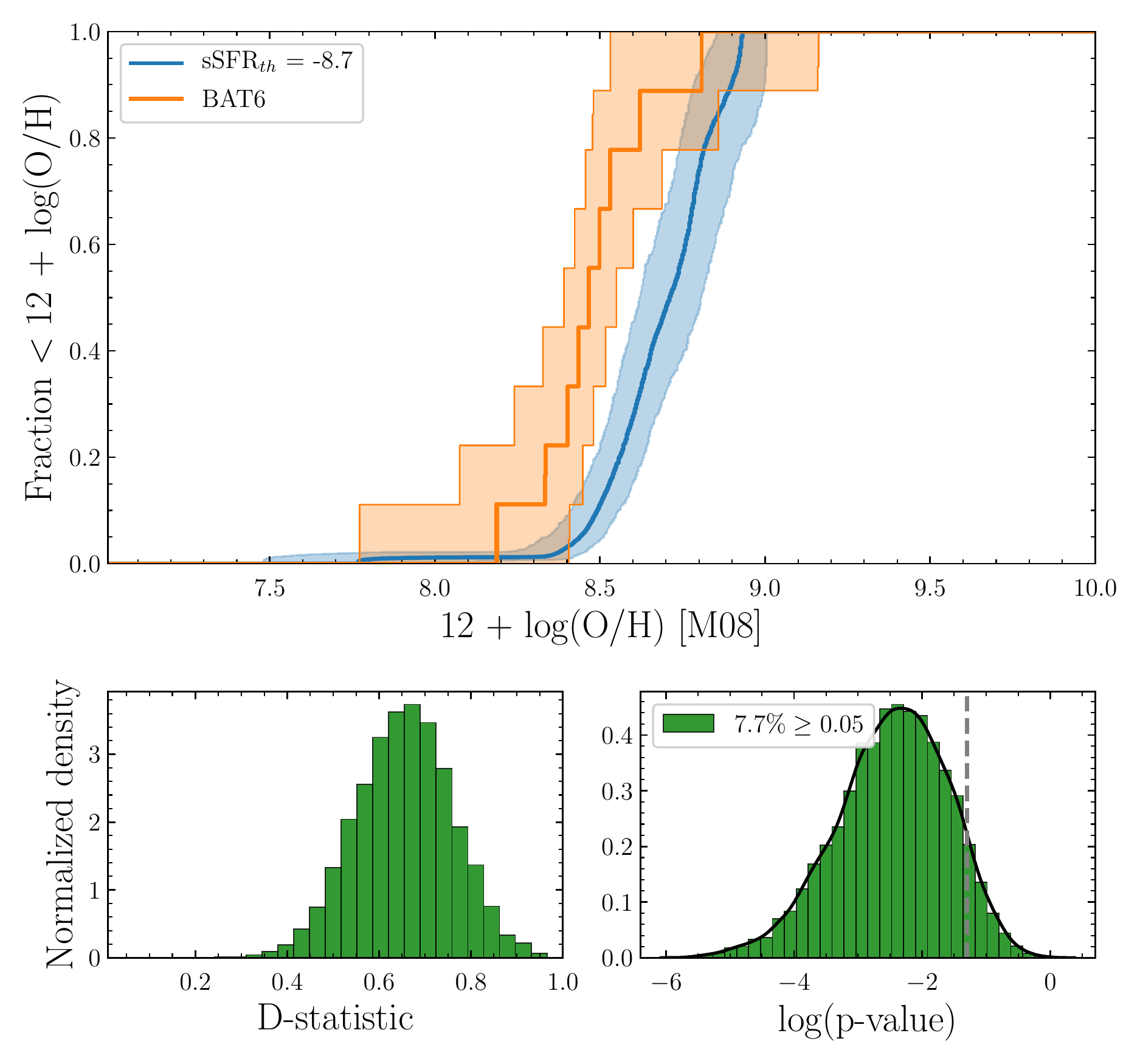}
\end{subfigure}
\caption{
\small
The result of the same analysis as presented in Section~\ref{sec:compar_samples} except using a sSFR cut on the MOSDEF sample of log(sSFR [yr$^{-1}$]) = -8.7.
}
\label{fig:all_CDF_MOSDEF_ssfr_cut}
\end{figure*}

To verify that enhanced star formation is not the main driving factor affecting the LGRB efficiency, we perform the same analysis above, but applying a cut only on the sSFR this time. As shown in Fig.~\ref{fig:all_CDF_MOSDEF_ssfr_cut}, even a sSFR cut of MOSDEF galaxies at $\rm{log(sSFR)}\geq-8.7$, (comparable with the sSFR of LGRB host galaxies) is not able to reconcile the stellar mass and metallicity distributions.

In general, we cannot exclude that a preference for galaxies with enhanced star formation (or starbursts) is also at play, but we can affirm that this is not the major factor driving the LGRB efficiency (see also \citealt{Graham2017}). 
We tested also the effect of various sSFR cuts on top of a metallicity cut. The impact is very mild and results in a slightly better agreement of the distributions for a metallicity cut between \logOH = 8.55 and \logOH = 8.7.
In \cite{Kelly2014} a preference for LGRB to explode in more compact galaxies (smaller half-light radii, higher SFR density and stellar mass density) compared to the SDSS star-forming galaxies is found, in addition to the preference for low-metallicities. 
However, considering the redshift range and low stellar-masses of our study, a morphological analysis cannot be performed.

The results obtained can be interpreted in terms of the conditions necessary for a massive star undergoing a collapse to form an LGRB.
A high metallicity would create too much wind-loss in the final stages of the progenitor's life, causing a loss of angular momentum that is necessary for the formation of an ultra-relativistic jet.
However, the threshold we find (corresponding to $~0.7\,Z_{\sun}$ in the M08 scale) is higher than the $0.1-0.3\,Z_{\sun}$ metallicity upper limit values predicted by most single-star progenitor models (e.g. \citealt{Yoon2006,Woosley2006}). Some studies pointed out that the \cite{Kewley2002} photoionization models on which the M08 method is based may overestimate oxygen
abundances by 0.2-0.5\,dex compared to the metallicity derived
using the so-called direct T$_e$ method (see e.g., \citealt{Kennicutt2003,Yin2007}). On the other hand it should also be noted that
the oxygen abundances determined using temperatures derived
from collisional-excited lines could be underestimated by 0.2-
0.3\,dex (see e.g. \,\citealt{Lopez-Sanchez2012,Nicholls2012}). 
A way to accommodate single star progenitors models with environments characterised by the higher metallicity values found in our works is to invoke chemically homogeneous mixing with very rapid rotation \citep{Brott2011} and weak magnetic coupling \citep{Georgy2012,Martins2013}. In such cases LGRB could be produced also up to solar metallicities, but it is still not clear whether their rates would correspond to the LGRB observed rates.   

Another possibility to be considered is an enhancement of the [O/Fe] in LGRB host galaxies. Indeed oxygen overabundances have been found in young and/or starburst galaxies (e.g. \citealt{Vink2000} and references therein; \citealt{Izotov2006}) due to the longer time scale needed to produce type Ia SNe, that are the main producer of iron, compared to type II SNe where oxygen is produced. This was also pointed out by \cite{Steidel2016} as an explanation of the higher stellar metallicity compared to the nebular one found for galaxies at $z>2$. Indications of low iron abundances compared to oxygen have been found by \cite{Hashimoto2018} in the host galaxies of two very low-redshift LGRBs: GRB\,980425 and GRB\,080517.
At the Z values we find, iron is the main driver of the wind mass-loss of WR stars \citep{Vink2005}. If the LGRB environment is characterised by oxygen overabundance, a [O/Fe]$\gtrsim0.5$ would imply iron metallicities in agreement with most single star LGRB models. 

Binary channels where the progenitor star is tidally spun-up by its companion \citep{deMink2009,Podsiadlowski2010} must also be considered.
The evolution of massive stars in binaries is more complex to model than as single stars (e.g. \citealt{Fryer2005,Yoon2015}). 
A few studies on evolutionary models of binary stars have started to investigate the effects of rotation and metallicity (e.g. \citealt{deMink2009,Eldridge2017}). 
In \cite{Song2016} the evolution of single and close binary stellar models (before any mass transfer) with strong core-envelope coupling is compared. Rotating massive stars in binary systems do not significantly lose their surface velocity, independent of the metallicity. 
Interestingly, the surface velocity increases with the initial stellar mass and the metallicity, and homogeneous evolution is more favoured at metallicities 
$Z\gtrsim0.5\,Z_{\odot}$ than at lower metallicities. The avoidance of the Roche lobe overflow phase during the main sequence phase is favoured in high-mass star models at metallicities $Z\lesssim0.5\,Z_{\odot}$. In the proposed scenario the primary star can enter the WR phase at an early stage of its evolution keeping fast rotation and high angular momentum. Even if the final stages of this evolution still need to be studied, this could be a channel for the formation of LGRBs also at moderately high metallicity.

More in general, it must be pointed out that the effect of metallicity goes beyond the final stages of the progenitor's life, and could also possibly affect the IMF of stars. The universality of the IMF is still debated, and different works pointed out the possibility of a metallicity dependence of the IMF, where a larger fraction of massive stars is produced at lower metallicity (e.g. \citealt{Marks2012,MartinNavarro2015}).

It is worth noting that the metallicities derived in this paper are integrated over the entire galaxy.
The possibility that the LGRB production site is situated in a low-metallicity pocket of a higher metallicity host should be considered. While this can not be excluded, various authors have shown that LGRB hosts are small and compact \citep{Lyman2017}, and when possible to resolve, little metallicity variation is found throughout the hosts (\citealt{Levesque2011,Kruehler2017,Izzo2017}; see however \citealt{Niino2017,Bignone2017} for considerations on metallicity variations).
We stress also that we used a simple step-function for the metallicity threshold because our small statistics do not allow us to constrain the shape of this function, however, in reality, it is more likely to be a smooth function of decreasing probability of hosting an LGRB with increasing metallicity.

Based on the fact that the hosts of the BAT6 LGRB sample represent a statistically complete sample of LGRB hosts, we can estimate the fraction of super-solar metallicity hosts (in the M08 scale).
With the conservative assumption that hosts without a metallicity measurement are super-solar (very unlikely, as they are mostly low mass galaxies), that fraction is less than 31$\pm15$\% at $z < 1$ and 33$\pm13$\% at $1 < z < 2$ (15$\pm15$\% and 13$\pm13$\%, respectively, if the host without metallicity measurement are sub-solar).

\section{Conclusions}\label{sec:conclu}

Using a complete and unbiased sample, we showed that the properties of LGRB host galaxies evolve between $z < 1$ and \ozt.
Their median stellar mass increases from  $\langle$ \logM $\rangle$ = 9.0$^{+0.1}_{-0.2}$ to 9.4$^{+0.2}_{-0.3}$, their median star formation rate increases from $\langle$~SFR~$\rangle$ = 1.3$^{+0.9}_{-0.7}$ to 24$^{+24}_{-14}$~\Msunyr, while their median metallicity remains constant at $\langle$~\logOH~$\rangle$ $\sim$ 8.45$^{+0.1}_{-0.1}$.
Based on the SF galaxy relation between SFR and stellar mass, the stellar mass evolution we found for LGRB host galaxies is weaker than that expected following their SFR evolution. If LGRB prefer to explode in environments for which the metallicity is below a certain threshold, such a (weaker) evolution is expected. In fact a fixed metallicity threshold would stifle LGRBs from exploding in high stellar mass galaxies, and at the same time would correspond to a higher stellar mass at higher redshift as the mass-metallicity relation evolves towards lower metallicities at fixed mass, or equivalently higher mass at fixed metallicity.

While performing the analysis of LGRB host galaxy properties, we revised some stellar mass values reported in the literature with proper SED fitting, confirming that the use of NIR photometry only can lead to overestimations of the stellar masses.
We looked at the LGRB FMR with the revised stellar masses, showing that there is still a shift with respect to the relation found by \citet{Mannucci2011}, at lower $\mu$, but that our sample is consistent with the MOSDEF star-forming galaxy sample.
This could be due to an underestimation of the FMR slope at low $\mu$ or to the current systematic uncertainties regarding evolution of metallicity calibrations with redshift.

We tested the hypothesis that LGRBs are \textit{pure} tracers of star formation (i.e., the probability of forming an LGRB is proportional to the SFR) by comparing the cumulative distributions of stellar mass, SFR, sSFR and metallicity of our sample with the ones of the COSOMOS2015UD (excluding metallicity) and MOSDEF representative surveys of star-forming galaxies at $1<z<2$.
Even if there is evidence for a preference of LGRB to explode in galaxies with enhanced star formation, we demonstrated that the major factor explaining the discrepancy between the mass and metallicity CDFs is a decrease of LGRB production in galaxies with metallicities above \logOH\ $\sim$ 8.55 in the M08 calibrator, although this threshold is to be cautiously treated as an indication rather than an absolute value due to statistics and calibrator robustness.
A lower LGRB production efficiency in higher metallicity environments can be understood in terms of the conditions necessary for the progenitor star to form a LGRB. The values found in this study invoke peculiar conditions of massive single star evolutionary models, and may be in better agreement with evolution in binary systems.

If this metallicity threshold is the only factor regulating the LGRB production efficiency, we expect LGRB to trace star formation in an unbiased manner once the bulk of the star-forming population of field galaxies is below this threshold. 
Assuming a threshold value of $Z_{\rm th}=0.7\,Z_{\odot}$, following the prescription of \cite{Langer2006}, and assuming that the LGRB luminosity function and density do not vary with redshift, this will happen for $z>3$. 
This scenario is in agreement with the findings of  
\cite{Greiner2015} and \cite{Perley2016}.
It is also supported by the decrease towards $z\sim3$ of the discrepancy of the stellar mass of the LGRB hosts and that of star-forming galaxies in surveys weighted by SFR. 
The collection of larger sample of high-$z$ GRBs with future dedicated satellites as the {\it THESEUS} mission \citep{Amati2018} will provide a viable way to probe the star formation history up to $z=10$ and beyond.

\section*{Acknowledgements}
This work is part of the BEaPro project (P.I.: SDV), funded by the French National Research Agency (ANR) under contract ANR-16-CE31-0003. JTP, SDV, RS, SB and ELF acknowledge the ANR support. 
This work benefits also of the support of the Programme National de Cosmologie et Galaxies (PNCG). 
JTP wishes to thank J.K Krogager and T. Charnock for fruitful discussions and C. Laigle for providing the mass completeness of the COSMOS2015 Ultra Deep survey.
SDV thanks Fabrice Martins, George Meynet and Fabian Schneider for very useful discussions. We thank T. Krühler for the data reduction of GROND observations and more in general for his openness in collaborating.
R.L.S. was supported by a UCLA Graduate Division Dissertation Year Fellowship, and also acknowledges a NASA contract supporting the ``WFIRST Extragalactic Potential Observations (EXPO) Science Investigation Team" (15-WFIRST15-0004), administered by GSFC.
JJ acknowledges support from NOVA and NWO-FAPESP grant for advanced instrumentation in astronomy.
AVG acknowledges support from the ERC via two Advanced Grants under grant agreements number 321323-NEOGAL and number 742719-MIST.
DC acknowledges support by the Centre National d’Etudes Spatiales and support by the Région Provence-Alpes-Côte d’Azur for the funding of his PhD.
This work is based in part on observations made with the \textit{Spitzer} Space Telescope, which is operated by the Jet Propulsion Laboratory, California Institute of Technology under a contract with NASA. Support for this work was provided by NASA through an award issued by JPL/Caltech.
Part of the funding for GROND (both hardware as well as personnel) was generously granted from the Leibniz-Prize to Prof. G. Hasinger (DFG grant HA 1850/28-1).
This research has made use of Astropy, a community-developed core Python package for Astronomy (Astropy Collaboration, 2013).
We acknowledge the use of Jochen Greiner's GRB website (http://www.mpe.mpg.de/~jcg/grbgen.html).

\bibliographystyle{aa} 
\bibliography{palmerio_biblio}

\appendix

\section{LGRB host galaxies: magnitudes, emission line fluxes and SEDs}\label{app:sampleBAT6}

\subsection*{GRB~061007 host: GROND magnitudes}
The host of GRB\,061007 was observed in the {\it griz} filters with the GROND instrument \citep{Greiner2008}. The data were reduced as outlined in \citet{Kruhler2008}. Photometric zero-points were obtained from GROND observations of SDSS fields taken right after the GRB field (see e.g. \citealt{Kruhler2011}). Photometry was measured with \textsc{SExtractor} (v2.8.6, \citealt{Bertin1996}). Final errors include both statistical errors and the uncertainties in photometric calibration.

\subsection*{GRB~100615A host: GROND and \textit{HST} magnitudes}

The host of GRB\,100615A was observed with the GROND instrument \citep{Greiner2008}. The data obtained with the {\it g,i,z} filters were reduced as outlined in \citet{Kruhler2008}. Photometric zero-points were obtained from GROND observations of SDSS fields taken right after the GRB field (see e.g. \citealt{Kruhler2011}). Photometry was measured with \textsc{SExtractor} (v2.8., \citealt{Bertin1996}). Final errors include both statistical errors and the uncertainties in photometric calibration.

{\it HST}-WFC3 near-infrared imaging observations were obtained with the F160W filter on 2010 December 16 from 21:38:48 UT to 22:01:01 UT (P.I.: A. Levan), for a total exposure time of 1.2 ks. We retrieved the resulting preview image from the MAST archive. Aperture photometry was made with the PHOTOM software part of the STARLINK\footnote{\url{http://starlink.eao.hawaii.edu/starlink}} package and calibrated using the standard WFC3 zeropoints
\footnote{\url{http://www.stsci.edu/hst/wfc3/analysis/ir_phot_zpt}}.

\subsection*{GRB~090201 host}
GRB 090201 was observed by IRAC \citep{Fazio2004} on the \textit{Spitzer} Space Telescope \citep{Werner2004} as part of the extended sample of the Swift Galaxy Host Legacy survey \citep{Perley2016b}. 
We subtracted nearby sources to provide a clean extraction aperture and performed aperture photometry on the host galaxy, and converted the resulting luminosity into a stellar mass, using the methods of \citet{Perley2016}.

\begin{landscape}
\begin{table}
\renewcommand{\arraystretch}{1.3}
\begin{center}
\footnotesize
\begin{tabular}{lccccccccccc}
\hline
\hline

   Name 	& Redshift	&      U   			&    B   	 		&    V   	 		&     R   	 		&     I   			&    g   			&    r   			&      i   	 		&      z   			& Ref \\
\hline
GRB050318 	& 1.4436 	&    	    		&    	    		&    	    		& $>$27.00  		&    	    		&    	    		&    	    		&    	    		&    	    		& 2 \\
GRB050802 	& 1.7117 	&    	    		&    	    		&    	    		&    	    		&    	    		&    	    		&    	    		&    	    		&    	    		& - \\
GRB060306 	& 1.5597 	& 25.61$\pm$0.2		& 25.19$\pm$0.1		&    	    		& 24.3$\pm$0.08		&    	    		& 25.06$\pm$0.1		&    	    		&    	    		& 23.72$\pm$0.17	& 2,6\\
GRB060814 	& 1.9223 	&    	    		&    	    		& 23.08$\pm$0.1		& 22.99$\pm$0.1		&    	    		& 23.11$\pm$0.09	& 22.94$\pm$0.08	& 22.37$\pm$0.09	& 22.58$\pm$0.19	& 2,6\\
GRB060908 	& 1.8836 	&    	    		&    	    		&    	    		& 25.73$\pm$0.18	&    	    		&    	    		&    	    		&    	    		&    	    		& 2 \\
GRB061007 	& 1.2623 	&    	    		&    	    		&    	    		& 24.9$\pm$0.17		&    	    		& 24.77$\pm$0.16	& 24.6$\pm$0.19		& $>$24.7	 		& $>$23.9	 		& 1,2 \\
GRB061121 	& 1.3160 	& 22.4$\pm$0.1		& 22.59$\pm$0.05	&    	    		& 22.54$\pm$0.04	& 22.36$\pm$0.1		& 22.41$\pm$0.04	&    	    		&   	    		& 22.09$\pm$0.09	& 2,4,8 \\
GRB070306 	& 1.4965 	& 23.02$\pm$0.6		&    	    		&    	    		& 22.92$\pm$0.09	&    	    		& 22.79$\pm$0.09	& 23.00$\pm$0.1		& 22.75$\pm$0.14	& 22.8$\pm$0.19		& 6 \\
GRB071117 	& 1.3293 	&    	    		&    	    		&    	    		&    	    		&    	    		& 24.4$\pm$0.1		& 24.7$\pm$0.2		& 24.8$\pm$0.3		& $>$24.4	 		& 3 \\
GRB080413B	& 1.1012 	&    	    		&    	    		&    	    		&    	    		&    	    		& 25.41$\pm$0.25	& 25.2$\pm$0.21		& 24.79$\pm$0.26	& $>$24.28	 		& 9 \\
GRB080602 	& 1.8204 	&    	    		&    	    		&    	    		& 22.95$\pm$0.02	&    	    		& 23.25$\pm$0.07	&    	    		& 22.98$\pm$0.09	& 22.9$\pm$0.12		& 3,7 \\
GRB080605 	& 1.6408 	&    	    		&    	    		&    	    		&    	    		&    	    		& 22.62$\pm$0.08	& 22.46$\pm$0.08	& 22.54$\pm$0.09	& 22.56$\pm$0.11	& 5 \\
GRB090926B	& 1.2427 	& 23.59$\pm$0.13	&    	    		&    	    		&    	    		&    	    		& 23.22$\pm$0.08	& 22.9$\pm$0.07		& 22.87$\pm$0.12	& 22.43$\pm$0.10	& 5 \\
GRB091208B	& 1.0633 	&    	    		&    	    		&    	    		&    	    		&    	    		&    	    		&    	    		&    	    		&    	    		& - \\
GRB100615A	& 1.3979 	&    	    		&    	    		&    	    		& 25.09$^{\dagger}\pm$0.03	&    	    		& 25.61$\pm$0.33	&    	    		& 24.73$\pm$0.26	& 24.03$\pm$0.23	& 1,10 \\
\hline
\hline
\end{tabular}
\end{center}
\caption{Observed optical AB magnitudes for the hosts of BAT6 at 1 < z < 2.
References: 1) this work; 2) \citet{Hjorth2012}; 3) \cite{Vergani2017}; 4) \citet{Perley2016}; 5) \cite{Kruhler2011}; 6) \citet{Perley2013}; 7) \citet{Rossi2012}; 8) \citet{Perley2015}; 9) \citet{Filgas2011}; 10) \citet{Blanchard2016}.\\
$\dagger$ F606W filter of \textit{HST}.
}

\label{tab:photometry_opt}
\end{table}

\begin{table}
\renewcommand{\arraystretch}{1.1}
\begin{center}
\footnotesize
\begin{tabular}{lccccccc}
\hline
\hline
   Name 	& Redshift	&     J   	 		&      H   	 		&     K$_s$	 		&   IRAC1	 		&    IRAC2	 		& Ref\\
\hline
GRB050318 	& 1.4436 	&   	    		&    	    		& $>$24.16  		& $>$25.22   		&    	     		& 2,4\\
GRB050802 	& 1.7117 	&   	    		&    	    		&    	    		& 24.86$\pm$0.29	&    	     		& 4\\
GRB060306 	& 1.5597 	&   	    		&    	    		& 21.94$\pm$0.1		& 21.54$\pm$0.02	& 21.32$\pm$0.09 	& 2,4,6\\
GRB060814 	& 1.9223 	& 22.09$\pm$0.15	& 21.88$\pm$0.1		& 21.88$\pm$0.1		& 21.43$\pm$0.04	& 21.18$\pm$0.09 	& 2,4,6\\
GRB060908 	& 1.8836 	&   	    		&    	    		& 24.38$\pm$0.43	& 24.73$\pm$0.20	&    	     		& 2,4\\
GRB061007 	& 1.2623 	&   	    		&    	    		& 22.9$\pm$0.37		& 23.71$\pm$0.15	&    	     		& 2,4\\
GRB061121 	& 1.3160 	& 22.05$\pm$0.35	&    	    		& 21.92$\pm$0.2		& 21.5$\pm$0.01		&    	     		& 2,4,8\\
GRB070306 	& 1.4965 	& 21.88$\pm$0.03	& 21.67$\pm$0.03	& 21.36$\pm$0.1		& 21.33$\pm$0.05	& 21.18$\pm$0.05 	& 2,4,6\\
GRB071117 	& 1.3293 	&   	    		&    	    		& 22.9$\pm$0.2		& $>$21.88	 		&    	     		& 3\\
GRB080413B	& 1.1012 	&   	    		&    	    		&    	    		& 23.22$\pm$0.09	&    	     		& 4\\
GRB080602 	& 1.8204 	& 22.39$\pm$0.32	& 21.75$\pm$0.29	& 22.55$\pm$0.05	& $>$22.8			&    	     		& 3,7\\
GRB080605 	& 1.6408 	& 22.16$\pm$0.3		& 21.96$\pm$0.05	& 21.75$\pm$0.3		& 21.61$\pm$0.1		&    	     		& 4,5\\
GRB090926B	& 1.2427 	& 21.81$\pm$0.13	& 21.84$\pm$0.26	& 21.39$\pm$0.19	& 21.42$\pm$0.04	&    	     		& 4,5\\
GRB091208B	& 1.0633 	&   	    		&    	    		&    	    		& $>$25.22    		&    	     		& 4\\
GRB100615A	& 1.3979 	&   	    		& 24.12$\pm$0.04	&    	    		& 23.84$\pm$0.14	&    	     		& 1,4\\
\hline
\hline
\end{tabular}
\end{center}
\caption{Observed near infrared AB magnitudes for the hosts of BAT6 at 1 < z < 2.
References: 1) this work; 2) \citet{Hjorth2012}; 3) \citet{Vergani2017}; 4) \citet{Perley2016}; 5) \cite{Kruhler2011}; 6) \citet{Perley2013}; 7) \citet{Rossi2012}; 8) \citet{Perley2015}.}
\label{tab:photometry_nir}
\end{table}

\end{landscape}

\begin{table*}
\renewcommand{\arraystretch}{1.1}
\small
\caption{
Stellar masses for the hosts of the BAT6 LGRB sample at $2<z<3$.
The galaxy stellar masses are computed using only the NIR \textit{Spitzer}/IRAC1 magnitudes or limits (\citealt{Perley2016}; see Sect.~\ref{subsec:prop_M*}).
References are : 1) this work; 2) \citet{Perley2016}.
}
\centering
\begin{tabular}{lccc}
\toprule
Name  & Redshift & $\mathrm{log(M_*/M_{\odot})}$ & $\mathrm{M_{ref}}$ \\
\midrule
070328 	& 2.0627 	& 10.0 	& 2	\\
090201 	& 2.1000 	& 10.9 	& 1	\\
100728B	& 2.106  	& <9.3 	& 2	\\
050922C	& 2.1995 	& <9.0 	& 2	\\
080804 	& 2.2059 	& 9.3  	& 2	\\
081221 	& 2.2590 	& 10.8 	& 2	\\
090812 	& 2.452  	& <9.4 	& 2	\\
081121 	& 2.512  	& 9.2  	& 2	\\
080721 	& 2.5914 	& <9.6 	& 2	\\
081222 	& 2.77   	& 9.6  	& 2	\\
050401 	& 2.8983 	& 9.6  	& 2	\\
\bottomrule
\end{tabular}
\label{tab:prop_sample_2z3}
\end{table*}

\begin{table*}
\renewcommand{\arraystretch}{1.1}
\begin{center}
\scriptsize
\caption{
Measured line fluxes in units of $10^{-17}$ \funit, corrected for Galactic foreground extinction.\\
References: 1) \citet{Kruhler2015}; 2) this work.
}
\label{tab:line_flux}
\begin{tabular}{lccccccccccc}
\hline
\hline
  Name 	    & Redshift	& \hbox{[O\,\textsc{ii}]\,$\lambda3726$} & \hbox{[O\,\textsc{ii}]\,$\lambda3729$}	& \hbox{[Ne\,\textsc{iii}]\,$\lambda3868$} &    \Hg		&    \Hb		& \hbox{[O\,\textsc{iii}]\,$\lambda4959$} & \hbox{[O\,\textsc{iii}]\,$\lambda5007$}&    \Ha		& \hbox{[N\,\textsc{ii}]\,$\lambda6583$} & Ref \\
\hline
GRB050318 	& 1.4436 	&    	    	&    	    		&    	    	&    	    	&    	    	&    	    	&    	    	&    		   	&    	    	& - \\
GRB050802 	& 1.7117 	&    	    	&    	    		&    	    	&    	    	&    	    	&    	    	&    	    	& 1.7$\pm$0.4	&    	    	& 2 \\
GRB060306 	& 1.5597 	& 0.7$\pm$0.4 	& 1.0$\pm$0.3 		& 1.1$\pm$1.4 	&    	    	& 1.4$\pm$2.0 	&    	    	& 3.3$\pm$4.6 	& 8.9$\pm$3.7 	& 2.4$\pm$1.4 	& 1 \\
GRB060814 	& 1.9223 	&    	    	& 26.3$\pm$3.7$^a$ 	&    	    	&    	    	& 8.3$\pm$3.4 	& 8.4$\pm$1.8 	& 31.$\pm$7.8 	& 28.0$\pm$5.7 	&    	    	& 1 \\
GRB060908 	& 1.8836 	&     	    	&    	    		&    	    	&    	    	&    	    	&    	    	&    	    	&    	    	&    	    	& 2 \\
GRB061007 	& 1.2623 	&     	     	& 2.4$\pm$0.3$^a$ 	& $<$2.0 	 	&    	    	& 1.0$\pm$0.4 	&     	     	& 9.5$\pm$1.4 	& 4.0$\pm$0.4 	& $<$2.4 		& 2 \\
GRB061121 	& 1.3160 	& 8.3$\pm$1.1 	& 18.4$\pm$1.1 		& 2.5$\pm$0.5 	& 4.2$\pm$1.4 	& 7.9$\pm$1.6 	& 7.9$\pm$1.6 	& 26.6$\pm$1.4 	& 40.0$\pm$0.9 	& 4.5$\pm$0.8 	& 2 \\
GRB070306 	& 1.4965 	& 9.1$\pm$0.7 	& 7.7$\pm$0.7 		& 1.9$\pm$0.4 	& 7.7$\pm$3.7 	& 11.6$\pm$1.4 	& 15.5$\pm$1.3 	& 46.0$\pm$3.6 	& 53.5$\pm$4.0 	& 6.4$\pm$0.4 	& 1 \\
GRB071117 	& 1.3293 	& 2.0$\pm$0.3 	& 3.4$\pm$0.3 		& $<$0.4 	  	&    	    	&    	    	& 3.0$\pm$0.6 	& 6.6$\pm$1.1 	& 5.6$\pm$1.0 	& $<$1.2 		& 2 \\
GRB080413B	& 1.1012 	& 0.6$\pm$0.2 	& 0.8$\pm$0.2 		& 0.2$\pm$0.2 	& 0.2$\pm$0.2 	&    	    	&    	    	& 2.8$\pm$0.9 	& 2.6$\pm$1.3 	&    	    	& 1 \\
GRB080602 	& 1.8204 	&     	    	& 28.$\pm$4.0$^a$ 	&    	    	&    	    	&    	    	&    	    	& 21.7$\pm$4.0	& 43.7$\pm$5.0	&    	    	& 2 \\
GRB080605 	& 1.6408 	& 7.9$\pm$1.1 	& 9.2$\pm$1.5 		&    	    	&    	    	& 7.7$\pm$1.5 	& 10.3$\pm$1.6 	& 29.6$\pm$4.6 	& 29.1$\pm$4.5 	& 4.0$\pm$0.7 	& 1 \\
GRB090926B	& 1.2427 	& 4.8$\pm$0.8 	& 7.1$\pm$0.8 		& $<$2.2 	 	& $<$2.8 	 	& 2.4$\pm$1.2 	& 3.1$\pm$1.0 	& 12.2$\pm$1.5 	& 11.5$\pm$1.2 	& $<$3.0 	 	& 2 \\
GRB091208B	& 1.0633 	&    	    	&    	    		&    	    	&    	    	&    	    	&    	    	&    	    	&    	    	&    	    	& - \\
GRB100615A	& 1.3979 	& 1.8$\pm$0.6 	& 2.7$\pm$0.6 		& 1.$\pm$0.3 	& $<$3.0 	  	& $<$2.8 	 	&    	    	&    	    	& 6.4$\pm$1.1 	& $<$1.6 		& 2 \\
\hline
\hline
\end{tabular}
\tablefoot{
$^a$ Cases where the \OII\, doublet is not resolved. The total integrated flux is reported in this column.}
\end{center}
\end{table*}

\begin{figure*}
 \begin{center}
 \includegraphics[width=0.6\linewidth]{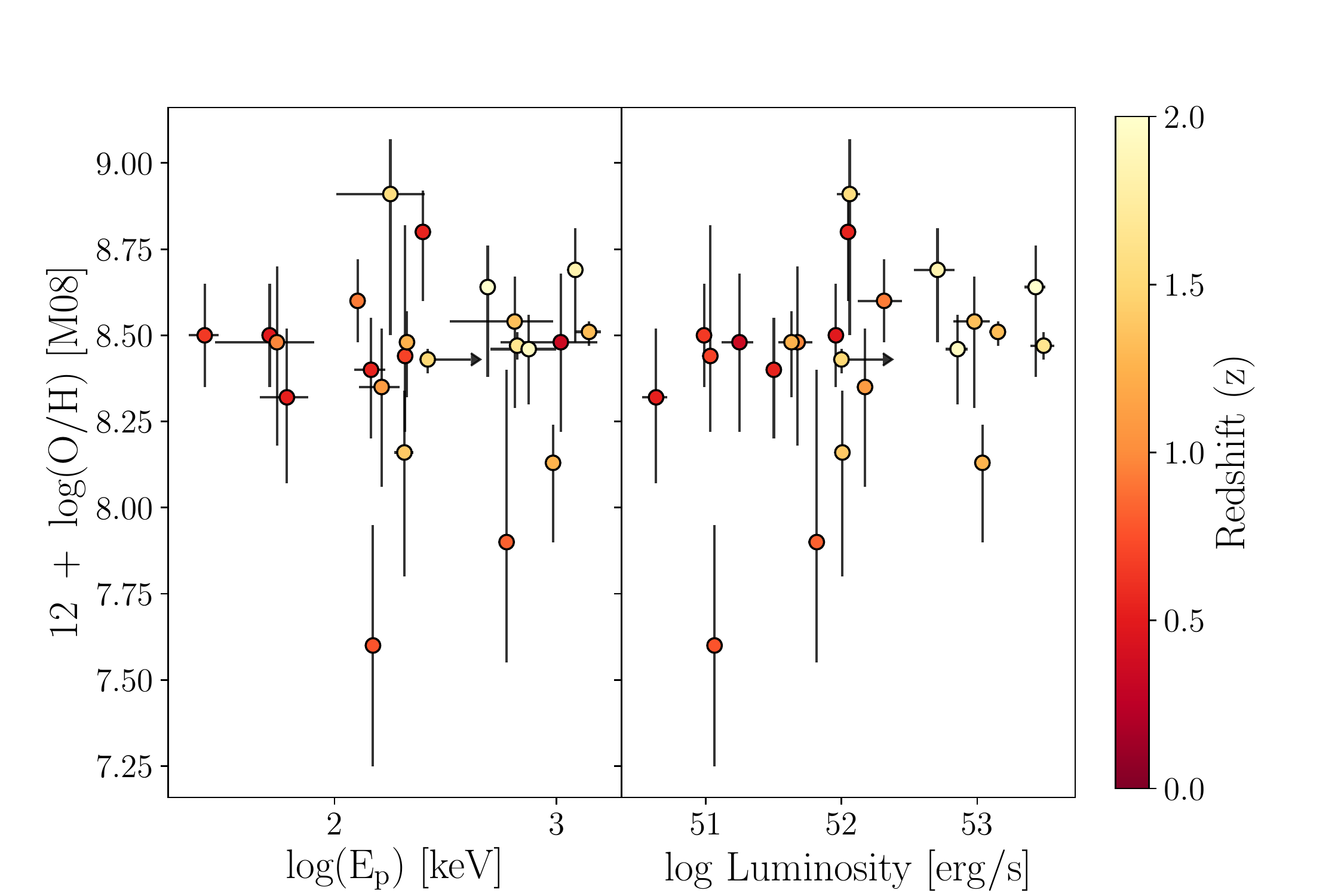}
\caption{
Metallicity of the LGRB host galaxies of the BAT6 sample at \ozt\  versus the peak of the $\nu\,F_{\nu}$ ({\it left panel}) and the isotropic-equivalent luminosity ({\it right panel}) of the prompt emission of the corresponding LGRB (from \citealt{Pescalli2016}). 
The points are colour-coded by redshift.
The arrows indicate lower limits.
}
\label{fig:prompt_correl}
\end{center}
\end{figure*}

\begin{figure*}
 \begin{center}
 \includegraphics[width=0.4\linewidth]{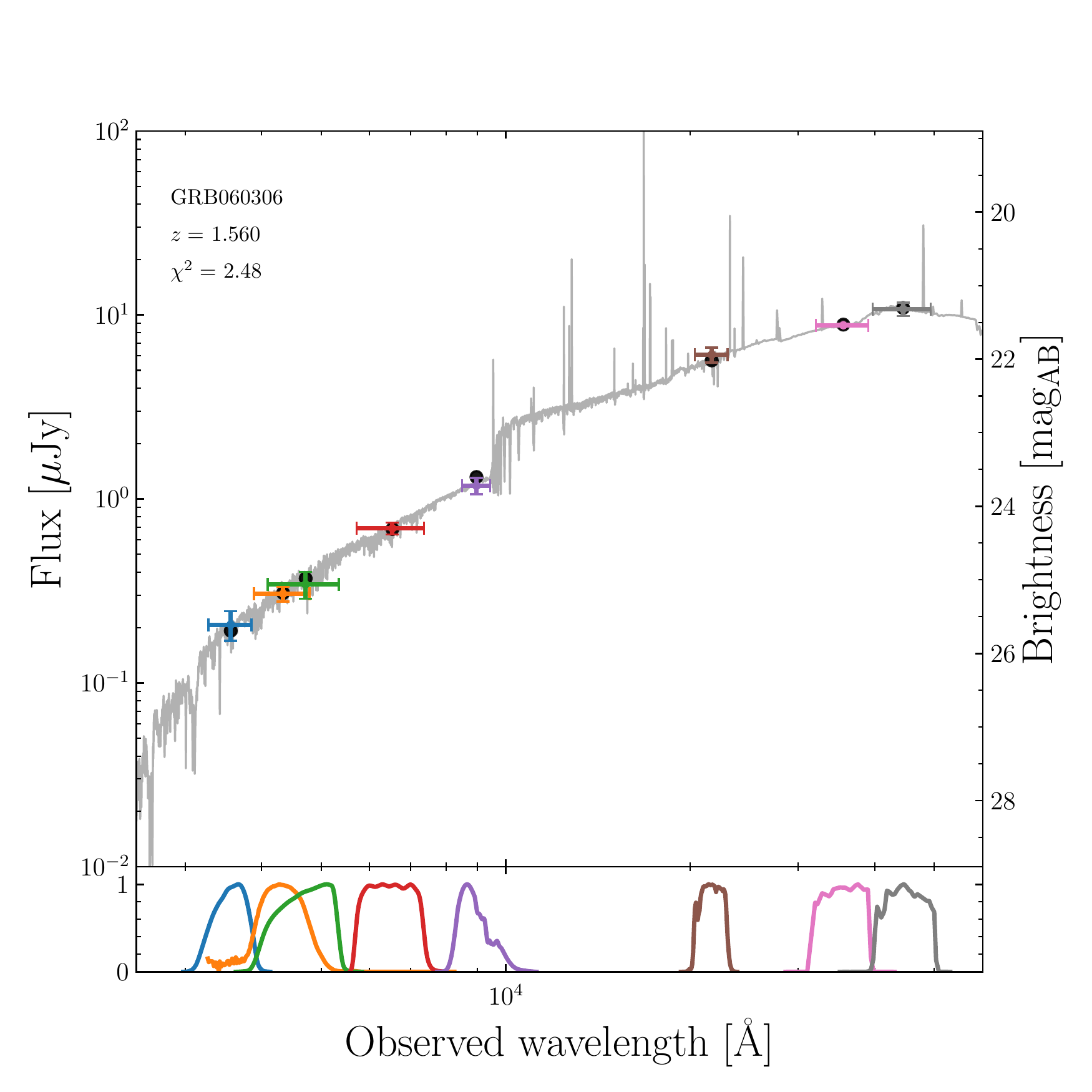}
 \includegraphics[width=0.4\linewidth]{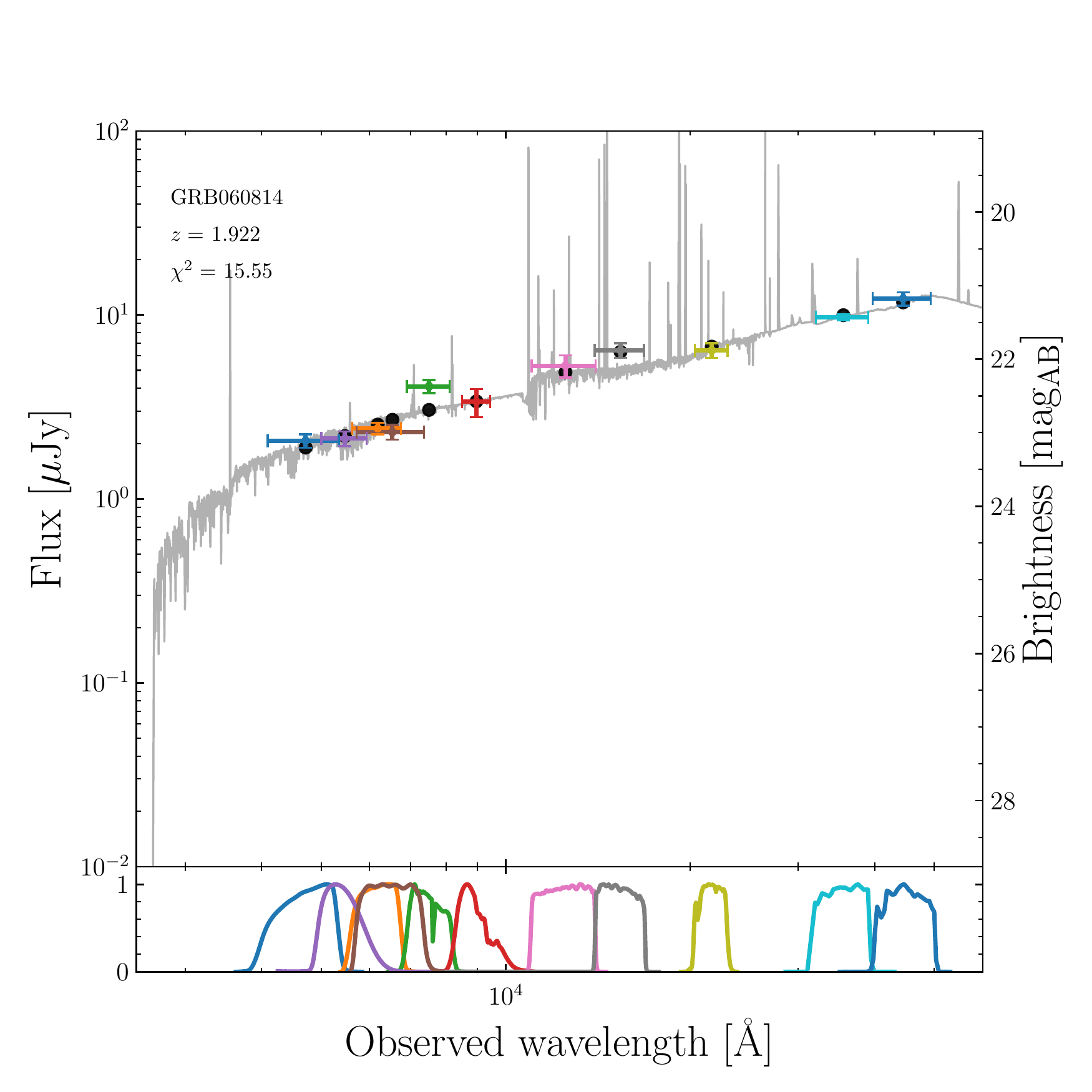}
 \includegraphics[width=0.4\linewidth]{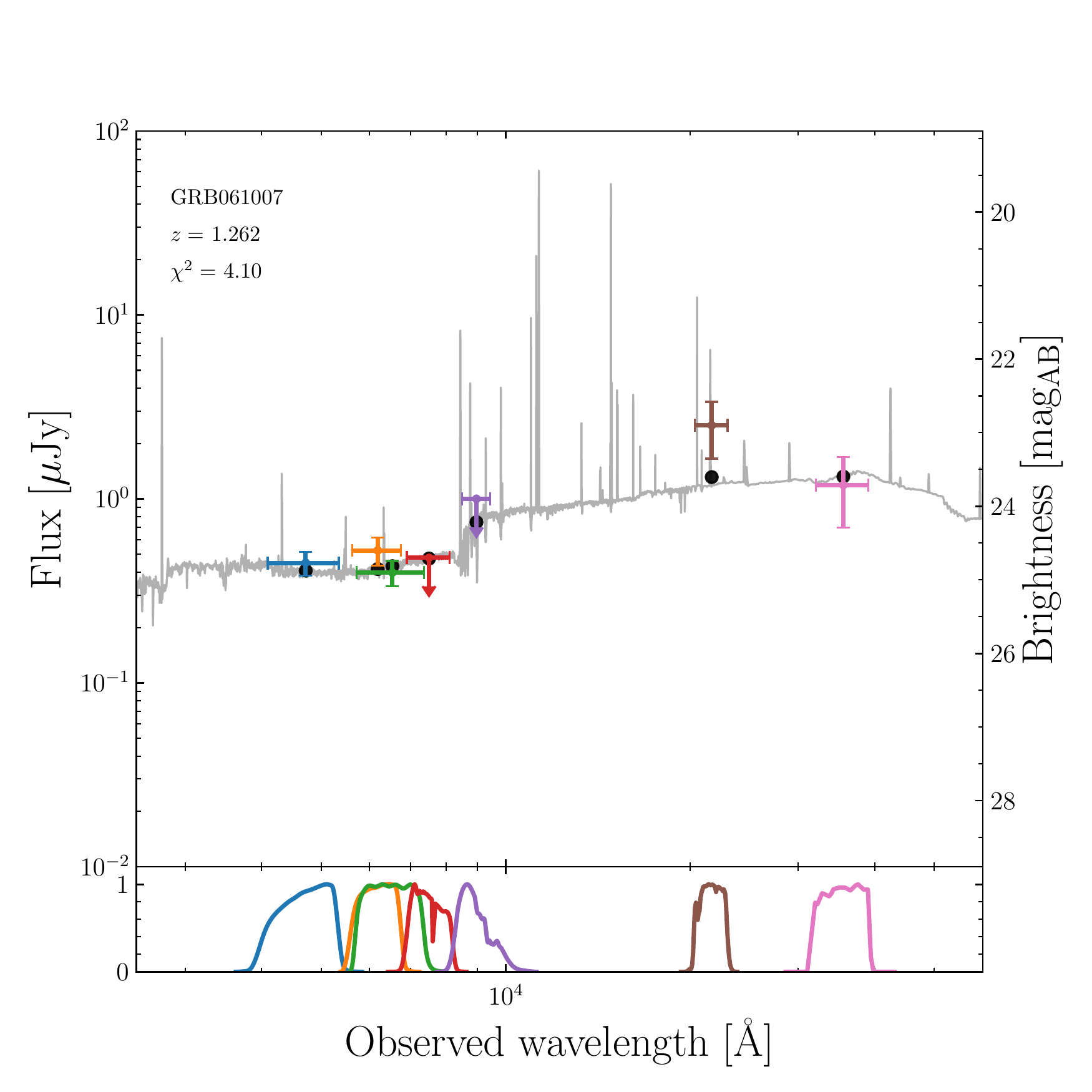}
 \includegraphics[width=0.4\linewidth]{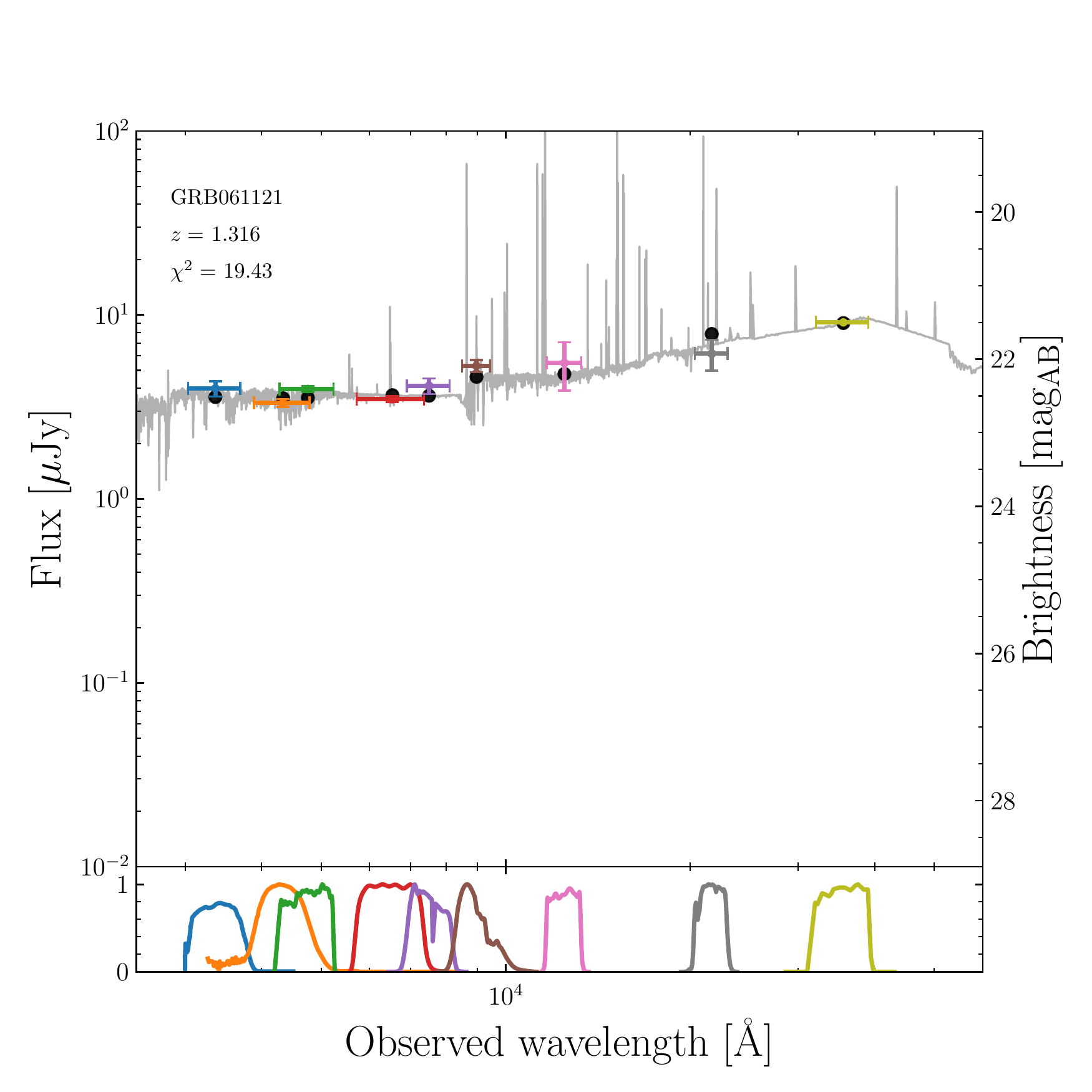}
 \includegraphics[width=0.4\linewidth]{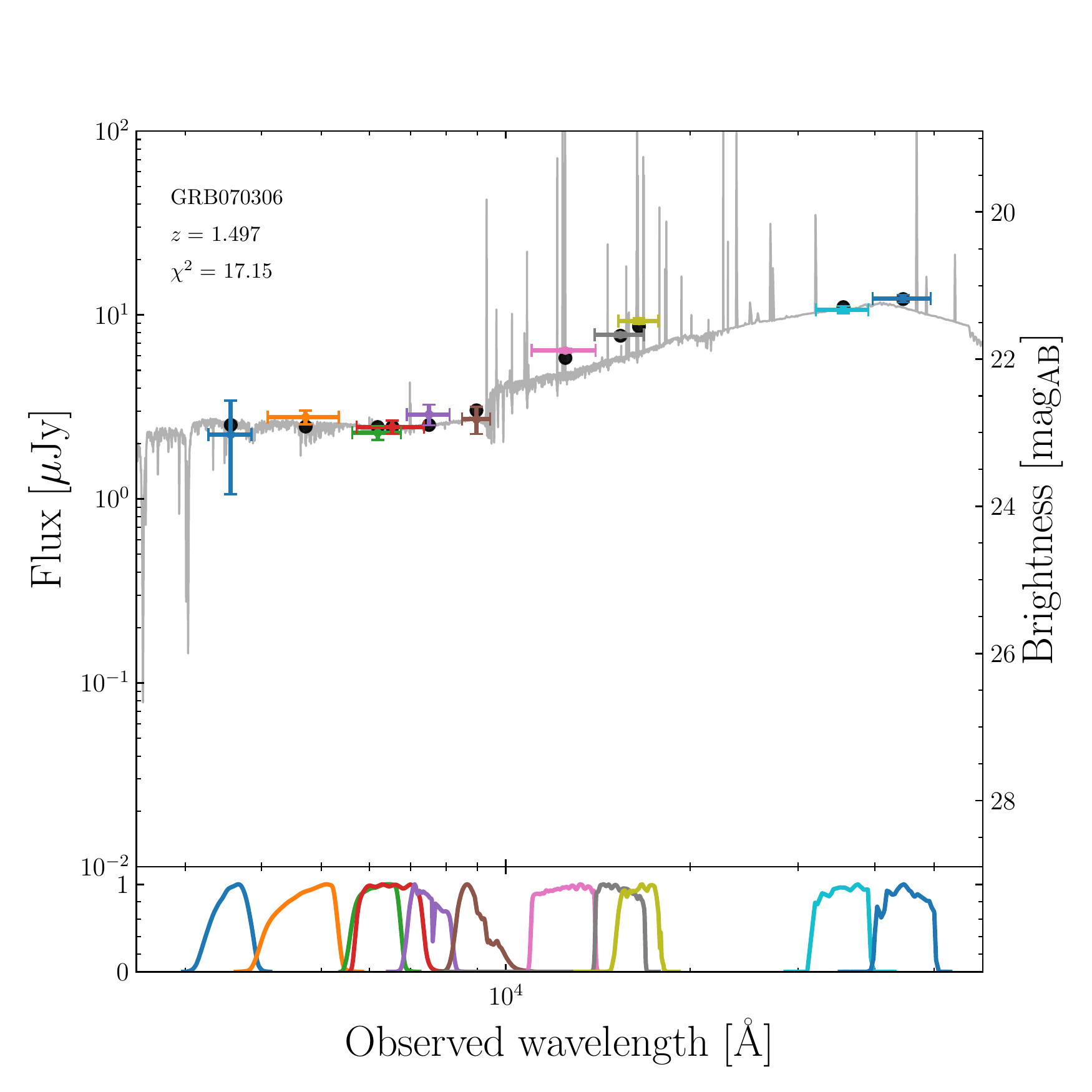}
 \includegraphics[width=0.4\linewidth]{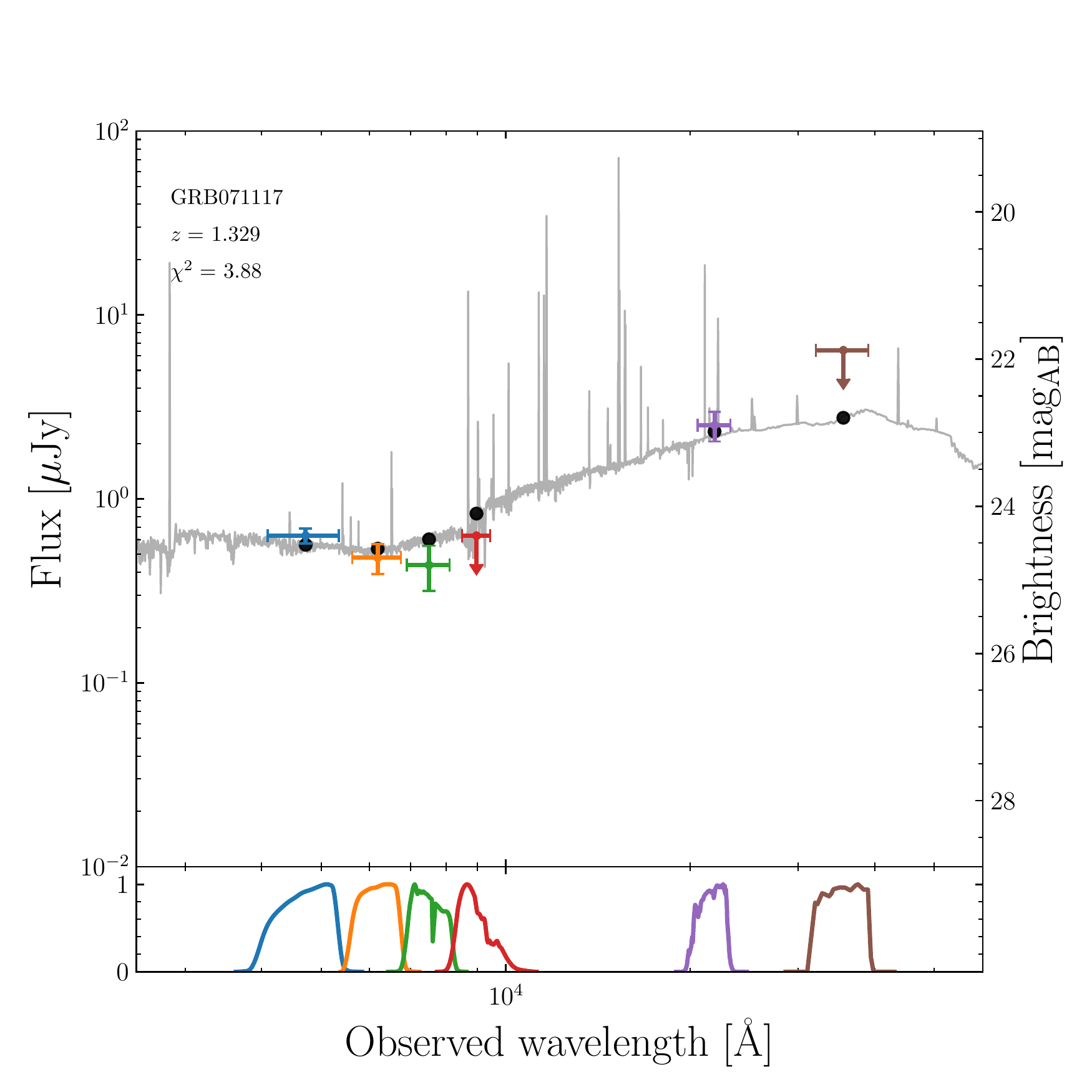}
\caption{
Best fit SEDs from \beagle\ are shown in grey, with the black circles representing the predicted filter values. Filter transmissions are shown in the bottom panels of each plot in the same colour as the corresponding observations shown as crosses in the upper panels. Upper limits are indicated by downward arrows.
The unreduced $\chi^2$ is shown in the top left of the upper panels.
}
\label{fig:best_fit_SED1}
\end{center}
\end{figure*}

\begin{figure*}
 \begin{center}

 \includegraphics[width=0.4\linewidth]{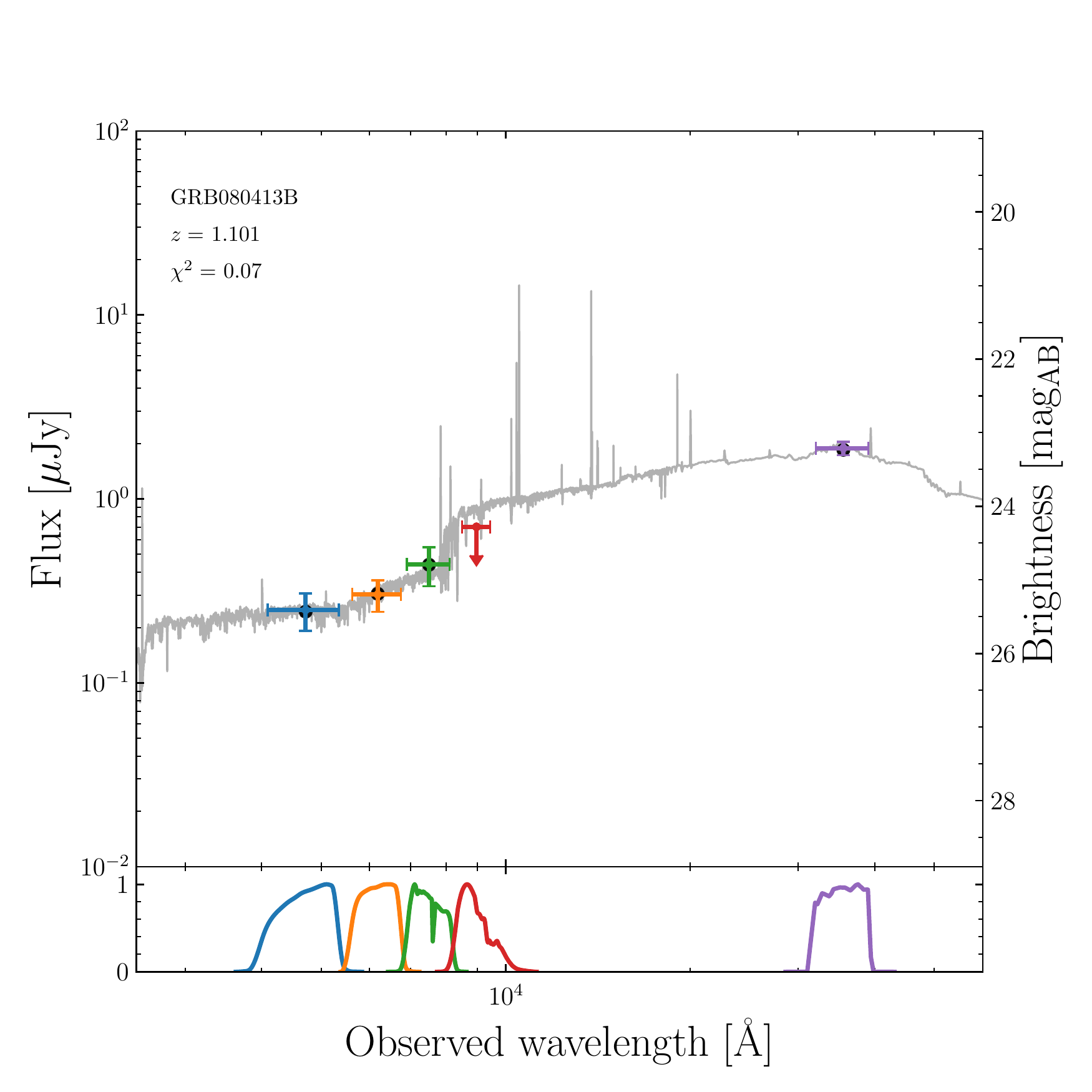}
 \includegraphics[width=0.4\linewidth]{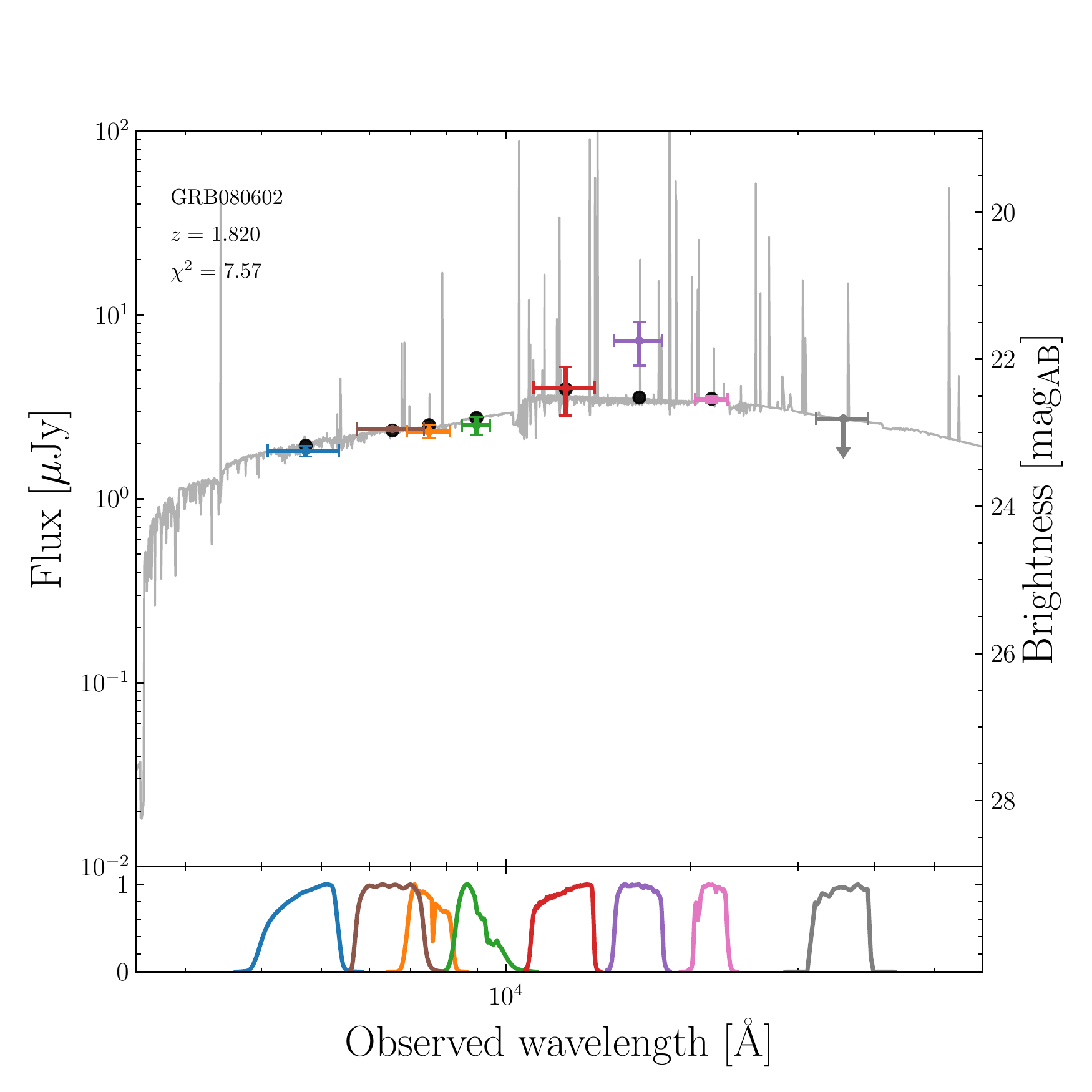}
 \includegraphics[width=0.4\linewidth]{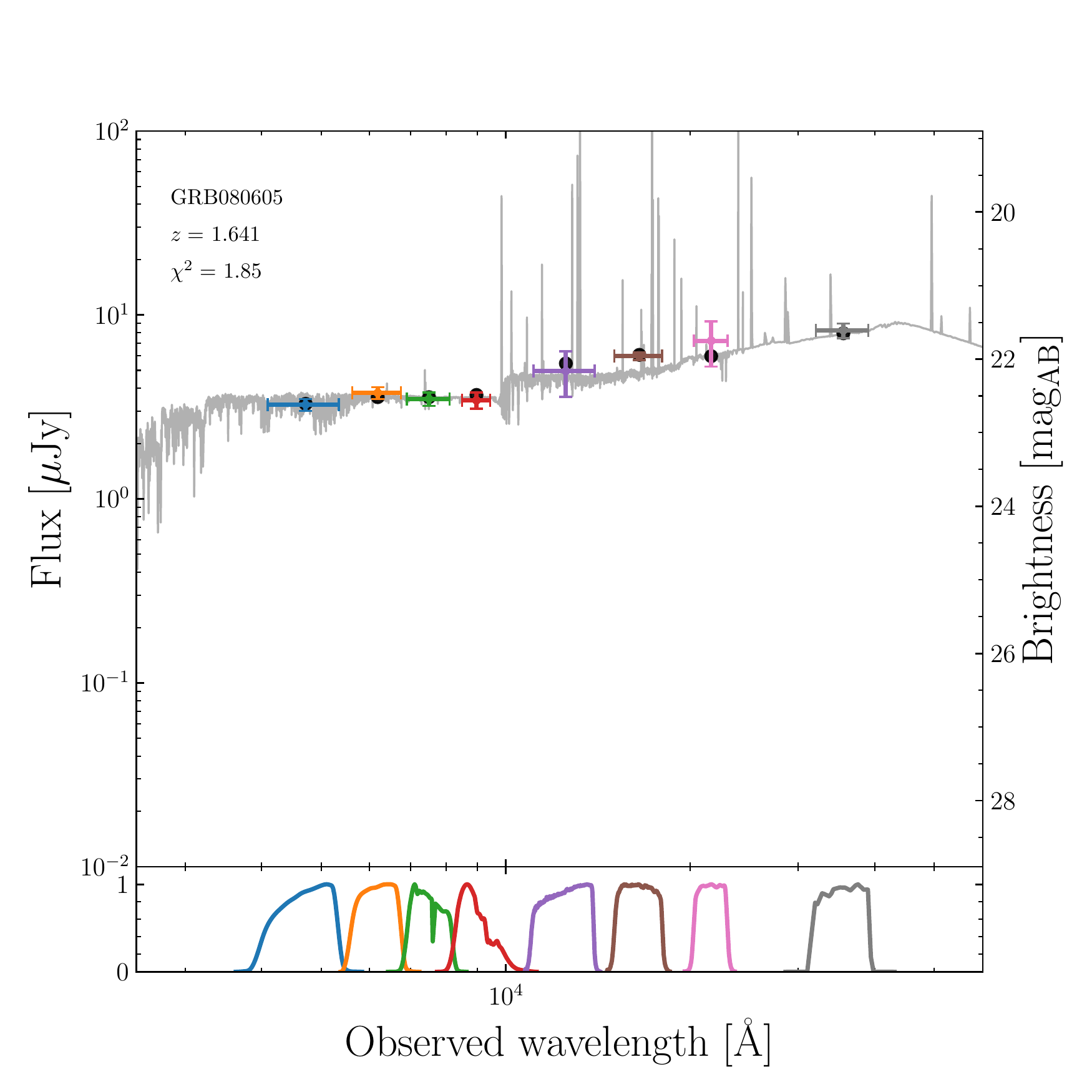}
 \includegraphics[width=0.4\linewidth]{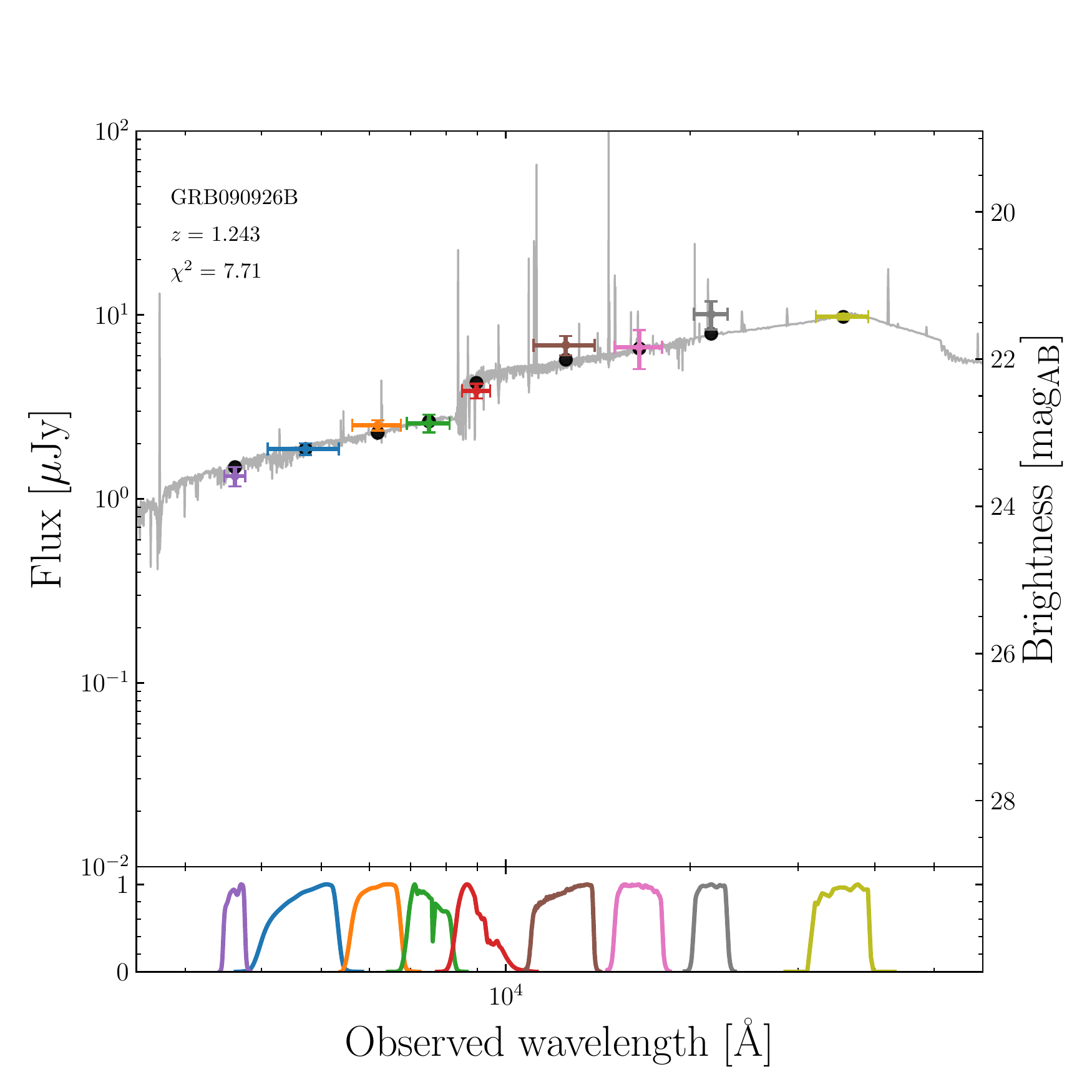}
 \includegraphics[width=0.4\linewidth]{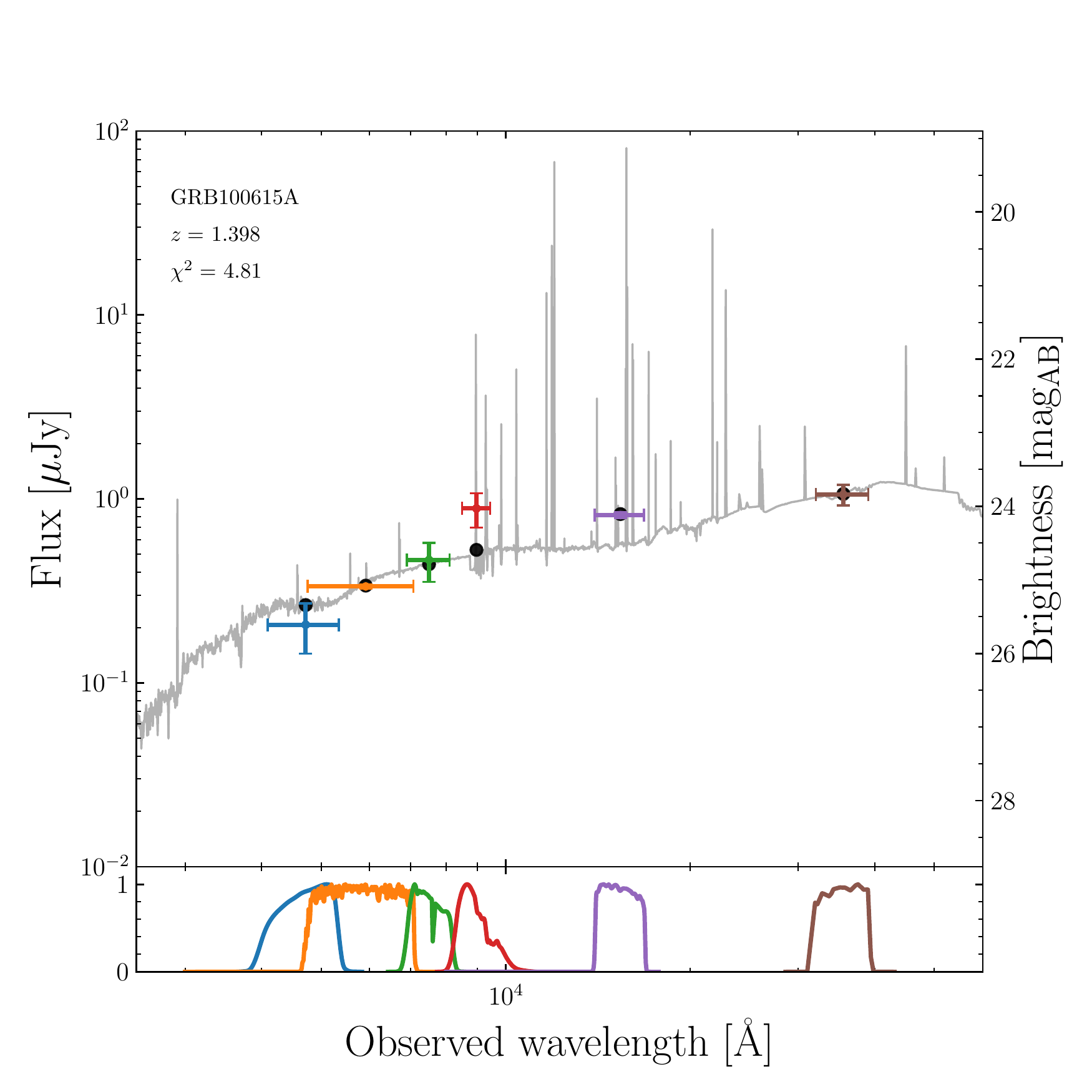}
\caption{
Same as Fig~\ref{fig:best_fit_SED1}
}
\label{fig:best_fit_SED2}
\end{center}
\end{figure*}

\end{document}